\documentclass[longauth]{aa}  
\usepackage{pifont}
\usepackage[colorlinks=true,allcolors=blue]{hyperref}

\usepackage{threeparttable}
\usepackage[rightcaption]{sidecap}
\usepackage{subcaption}

\usepackage{graphicx}
\usepackage{txfonts}
\newcommand{\cpurple}{}

\usepackage{xspace}
\newcommand{\kms}{km~s$^{-1}$\xspace}
\newcommand{\ergs}{erg~s$^{-1}$\xspace}

\newcommand{\hb}{H$\beta$\xspace}
\newcommand{\ha}{H$\alpha$\xspace}
\newcommand{\heii}{He\,{\sc{ii}}\xspace}

\newcommand{\feii}{Fe\,{\sc{ii}}\xspace}

\newcommand{\oi}{[O\,{\sc{i}}]\xspace}
\newcommand{\oiii}{[O\,{\sc{iii}}]\xspace}

\newcommand{\sii}{[S\,{\sc{ii}}]\xspace}

\newcommand{\civ}{C\,{\sc{iv}}\xspace}
\newcommand{\ciii}{C\,{\sc{iii}}]\xspace}
\newcommand{\cii}{[C\,{\sc{ii}}]\xspace}

\newcommand{\nii}{[N\,{\sc{ii}]}\xspace}

\newcommand{\hii}{H\,{\sc{ii}}\xspace}

\newcommand{\mgii}{Mg\,{\sc{ii}}\xspace}

\newcommand{\VDES}{VDES J0224--4711\xspace}

\newcommand{\HSC}{HSC J0859+0022\xspace}

\begin{document}

   \title{GA-NIFS: sRMS, a new method for disentangling resolved and unresolved emission in integral field spectroscopy.\\ Application to distant quasars}

   \titlerunning{Spatial RMS as a new method for IFS analysis of distant QSOs}

   \author{Michele Perna
          \inst{\ref{iCAB}}\thanks{e-mail: mperna@cab.inta-csic.es}
          \and
          Santiago Arribas\inst{\ref{iCAB}}
          \and 
          Carlota Prieto-Jim\'enez\inst{\ref{iCAB},\ref{iUCM}}
          \and 
          Isabella Lamperti\inst{\ref{iCAB}}
          \and
          Lorenzo Ulivi\inst{\ref{iCAB}}
          \and
          Hannah \"{U}bler\inst{\ref{iMPE}}
          \and
          Giovanni Cresci\inst{\ref{iOAA}}
          \and
          Bruno Rodr\'iguez~Del~Pino\inst{\ref{iCAB}}
          \and
          Torsten Böker\inst{\ref{iESAusa}}
          \and
          Andrew~J.~Bunker\inst{\ref{iOxf}}
          \and
          St\'ephane~Charlot\inst{\ref{iSor}}
          \and
          Roberto~Maiolino\inst{\ref{iKav},\ref{iCav},\ref{iUCL}}
          \and
          Chris~J.~Willott\inst{\ref{iNRC}}
          \and
          Ivan Delvecchio\inst{\ref{iOABo}}
          \and  
          Elena Bertola\inst{\ref{iOAA}}
          \and
          Madeline~A.~Marshall\inst{\ref{iLOS}}
          \and
          Eleonora~Parlanti\inst{\ref{iNorm}}
          \and
          Pablo~G.~P\'erez-Gonz\'alez\inst{\ref{iCAB}}
          \and
          Giacomo~Venturi\inst{\ref{iOAA}}
          \and
          Sandra~Zamora\inst{\ref{iNorm}}
          \and
          Filippo~Mannucci\inst{\ref{iOAA}}
          \and
          Quirino D'Amato\inst{\ref{iOARoma}}
          \and
          Alessandra De Rosa\inst{\ref{iOARoma}}
          \and 
          Ismael García-Bernete\inst{\ref{iCAB}}
          \and
          Alessandro Marconi\inst{\ref{iUNIFI},\ref{iOAA}}
          \and
          Miguel~Pereira-Santaella\inst{\ref{iIFF}}
          \and 
          Fabio~Rigamonti\inst{\ref{iOABrera}}
          \and
          Martina~Scialpi\inst{\ref{iOAA}}
          \and
          Paola Severgnini\inst{\ref{iOABrera}}
          \and
          Cristian Vignali\inst{\ref{iOABo},\ref{iUNIBo}}
          \and
          Maria Vittoria Zanchettin\inst{\ref{iOARoma}}
          }

   \authorrunning{M. Perna et al.}
   \institute{
            Centro de Astrobiolog\'ia (CAB), CSIC--INTA, Cra. de Ajalvir Km.~4, 28850 -- Torrej\'on de Ardoz, Madrid, Spain\label{iCAB}
    \and 
            Departamento de F\'{i}sica de la Tierra y Astrof\'{i}sica, Facultad de Ciencias F\'{i}sicas, Universidad Complutense de Madrid, E-28040, Madrid, Spain\label{iUCM}
    \and
            Max-Planck-Institut f\"ur extraterrestrische Physik (MPE), Gie{\ss}enbachstra{\ss}e 1, 85748 Garching, Germany\label{iMPE}
    \and 
            INAF - Osservatorio Astrofisico di Arcetri, Largo E. Fermi 5, I-50125 Firenze, Italy\label{iOAA}
    \and    
            European Space Agency, c/o STScI, 3700 San Martin Drive, Baltimore, MD 21218, USA\label{iESAusa}
    \and
            Department of Physics, University of Oxford, Denys Wilkinson Building, Keble Road, Oxford OX1 3RH, UK\label{iOxf}
    \and
            Sorbonne Universit\'e, CNRS, UMR 7095, Institut d’Astrophysique de Paris, 98 bis bd Arago, 75014 Paris, France\label{iSor} 
    \and
            Kavli Institute for Cosmology, University of Cambridge, Madingley Road, Cambridge, CB3 0HA, UK\label{iKav}
    \and
            Cavendish Laboratory - Astrophysics Group, University of Cambridge, 19 JJ Thomson Avenue, Cambridge, CB3 0HE, UK\label{iCav}
    \and
            Department of Physics and Astronomy, University College London, Gower Street, London WC1E 6BT, UK\label{iUCL}  
    \and
            NRC Herzberg, 5071 West Saanich Rd, Victoria, BC V9E 2E7, Canada\label{iNRC}
    \and 
            INAF – Osservatorio di Astrofisica e Scienza dello Spazio di Bologna, Via Gobetti 93/3, I-40129 Bologna, Italy\label{iOABo}
    \and
            Los Alamos National Laboratory, Los Alamos, NM 87545, USA\label{iLOS}
    \and
            Scuola Normale Superiore, Piazza dei Cavalieri 7, I-56126 Pisa, Italy\label{iNorm}
    \and
            INAF - Istituto di Astrofisica e Planetologia Spaziali, Via del Fosso del Cavaliere, 00133 Roma, Italy\label{iOARoma}
    \and
            Università di Firenze, Dipartimento di Fisica e Astronomia, via G. Sansone 1, 50019 Sesto F.no, Firenze, Italy\label{iUNIFI}
    \and
            Instituto de F\'isica Fundamental, CSIC, Calle Serrano 123, 28006 Madrid, Spain\label{iIFF}
    \and    
            INAF – Osservatorio Astronomico di Brera, via Brera 28, 20121 Milano, Italy\label{iOABrera}
    \and
            Dipartimento di Fisica e Astronomia ‘Augusto Righi’, Università degli Studi di Bologna, Via Gobetti 93/2, 40129 Bologna, Italy\label{iUNIBo}
             }



  \abstract
   {
   The interpretation of emission-line profiles in type 1 active galactic nuclei (AGNs) is often complicated by the dominance of bright, spatially unresolved nuclear continuum and broad-line region (BLR) emission over narrower and typically blended components from the narrow-line region (NLR) and the \hii regions of the host galaxy. This remains challenging even with integral-field spectroscopy (IFS), as the relevant spatial variations can occur on scales smaller than the angular resolution, particularly in the compact host of distant AGN, and are therefore not necessarily accessible to conventional spaxel-by-spaxel analyses.

   } 
   {
   We introduce the spatial root-mean-square (sRMS) technique, a new approach to analyse single-epoch IFS that uses the spatial variance of spectra extracted from partially overlapping apertures to isolate off-nuclear emission of high-redshift type 1 AGNs, and investigate whether the resulting kinematic information can improve the decomposition of their integrated spectra.
   }
   {
    We construct sRMS spectra from ensembles of apertures covering the unresolved nucleus. Emission that is spatially invariant on the scales probed by the apertures is consequently suppressed, whereas spatially varying emission is retained. Using realistic JWST/NIRSpec IFS simulations, we optimise the aperture geometry and test the sensitivity of the method to spatially extended and spatially offset emission. We then apply the optimised procedure to JWST/NIRSpec observations of two $z\sim6.5$ quasars. We model the resulting sRMS spectra with multiple Gaussian components to determine the kinematics of the spatially varying emission, and use these kinematic constraints to model the integrated spectra.
   }
   {{
   The sRMS technique identifies spatially varying narrow-line emission on scales substantially ($\sim5\times$) smaller than the NIRSpec point-spread function and provides robust constraints on the velocity offsets and widths of individual kinematic components. For the observed quasars, imposing these kinematic constraints on the integrated-spectrum fits reduces significantly the degeneracy of multi-component decompositions.} Simulations further demonstrate the potential of the method to detect spatially offset BLR emission from close dual type 1 AGNs, down to projected separations of $\sim200$ pc at $z\sim6.5$ for a BLR flux ratio of $\sim10$.
   }
   {}


   \maketitle
%

\section{Introduction}

Optical spectra of active galactic nuclei (AGNs) exhibit a characteristic set of components (e.g. \citealt{Seyfert1943, Osterbrock2006}). In particular, so-called type~1 AGNs are characterised by an ultraviolet-optical continuum typically dominated by a blue, non-stellar emission arising from the accretion disk surrounding the central supermassive black hole (SMBH); superimposed on this continuum are broad permitted emission lines (e.g. \ha, \hb), which originate in the high-velocity gas of the broad-line region (BLR) located at sub-parsec scales. Narrower emission lines, both permitted and forbidden (e.g. \oiii\!$\lambda\lambda$4960,5008, \nii\!$\lambda\lambda$6550,85, \sii\!$\lambda\lambda$6718,32), arise from the more extended and lower-density narrow-line region (NLR). In many cases, the spectra of type~1 AGNs also show prominent \feii emission complexes, believed to originate in dense gas at intermediate distances between the BLR and NLR \citep[e.g.][]{Kovacevic2010, Marinello2016}.

Reverberation mapping has leveraged multi-epoch spectroscopic monitoring campaigns to study the temporal variability of the spectra of type 1 AGNs (e.g. \citealt{Peterson1993, Shen2024}). These observations, typically spanning months to years, are used to trace temporal changes in the AGN continuum and the response of the BLR emission lines. The time lag between continuum and broad-line flux variations yields a characteristic size for the BLR. Assuming the gas motions are virialized, the black hole mass can then be estimated from the line width of the variable BLR emission and the inferred radius. 

To isolate the variable components in spectroscopic observations, \citet{Peterson1998a} introduced the use of the root-mean-square (RMS) spectrum, constructed  deriving the RMS of the individual-epoch spectra taken over time. The RMS spectrum, measuring the variations around the mean, emphasizes only the portions of the spectrum that vary over time, typically the broad emission lines and continuum, while suppressing constant features such as the narrow emission lines, galactic absorption lines, and non-variable continuum. As such, the width of the broad lines derived from the RMS spectrum more accurately reflects the true kinematics of the variable BLR gas, and is therefore preferred for virial mass estimates. This technique has become a cornerstone of modern reverberation mapping studies 
\citep[e.g.][]{Kaspi2000, Grier2017,Shen2019}.

Although reverberation mapping provides the most direct measurements of SMBH masses, the long monitoring campaigns required limit its applicability to relatively small samples (e.g. \citealt{Trevese2014}). Consequently, the vast majority of black hole masses are estimated from single-epoch spectra, where the BLR radius is inferred from the empirical radius--luminosity relation and the velocity dispersion is measured from the width of the broad emission lines \citep[e.g.][]{Kaspi2000}. The accuracy of these estimates therefore critically depends on a reliable modelling of the broad and narrow emission-line components in single-epoch spectra (e.g. \citealt{Vietri2020,Scholtz2026blr}). This task becomes increasingly challenging in luminous quasars (QSOs), where blending between strong BLR and \feii complexes, NLR emission associated with strong outflows (e.g. \citealt{Fiore2017, Venturi2026ganifs}), and underlying continuum can severely hamper the identification and modelling of the different spectral components.

\begin{figure*}
   \centering
   \includegraphics[width=0.99\textwidth, trim=0mm 1mm 0mm 5mm,clip]{{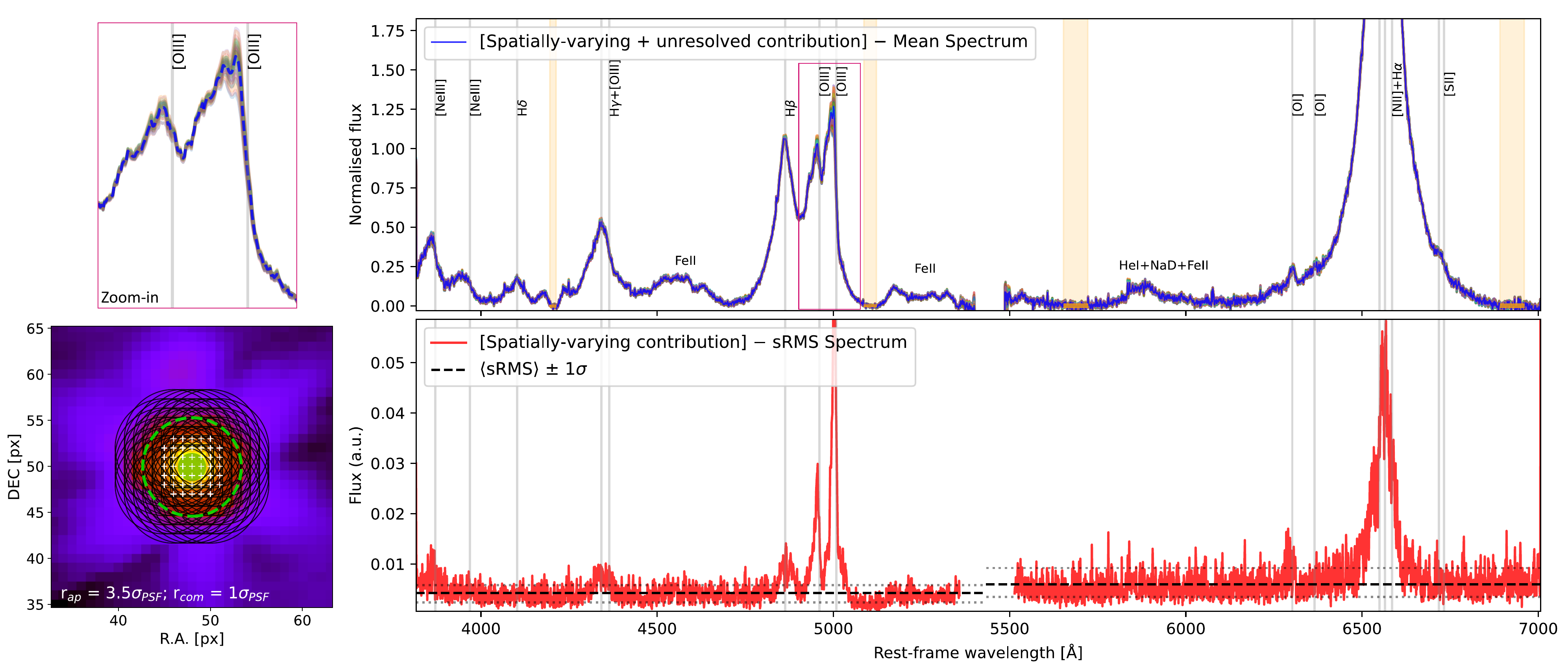}}

    \caption{  \VDES white light image, mean and sRMS spectra from JWST/NIRSpec IFS. {\it Bottom left:} The black circles indicate the individual extraction apertures used to obtain the ensemble of (normalised and continuum-subtracted) spectra entering the sRMS analysis, {\cpurple the white crosses mark the centre of the apertures}. The green dashed circle marks the reference aperture centred on the QSO position, for an aperture radius $r_{\rm ap} = 3.5\sigma_{\rm PSF}$, while the filled green circle identifies the region common to all apertures, for $r_{\rm com} = \sigma_{\rm PSF}$ (Sect.~\ref{sec:technique}). {\it Right:} 
    The upper panel shows the mean spectrum (thick blue line) together with the individual aperture spectra (coloured lines), which overlap with the mean except for small portions (e.g. around \oiii, {\cpurple see zoom-in inset on the top-left, and} Fig.~\ref{fig:VDESmeanzoomin}).
    The orange areas indicate the regions used for continuum modelling and subtraction. The lower panel presents the corresponding sRMS spectrum, which highlights the spatially varying emission associated with the narrow emission lines \hb, \oiii, \oi, \ha, and \nii, while largely suppressing the {\cpurple spatially unresolved emission from the AGN nucleus}. The horizontal dashed line indicates the average {\cpurple zero-}level of the sRMS spectrum, arising from residual variations introduced by noise and the wavelength dependence of the PSF; the dotted lines mark the $\pm 1\sigma$ scatter around this level. The {\cpurple zero-}level is systematically higher in the second NIRSpec detector than in the first because of its higher noise. }
    \label{fig:figure1}%
    \end{figure*}

Inspired by the temporal RMS approach, we explore in this work a new method based on the RMS spectrum in the spatial domain, enabled by integral field spectroscopy (IFS). Instead of combining spectra obtained at different epochs, we construct a spatial RMS (hereafter sRMS) spectrum from the ensemble of spectra extracted at different spatial positions within a single 
IFS data cube, all with an overlapping region centred on the AGN (see Fig.~\ref{fig:figure1}). The sRMS spectrum highlights spectral components that vary across {\cpurple the sampled area}, such as extended emission-line regions, ionised outflows, compact stellar clumps, or AGN ionisation cones, while suppressing {\cpurple the brighter} emission that is spatially invariant over the scales probed, namely the unresolved AGN continuum, BLR emission from hydrogen Balmer transitions, and \feii emission. In this sense, the sRMS spectrum acts as the spatial counterpart of the classical temporal RMS spectrum: whereas the latter isolates compact, time-variable emission from the BLR, the former identifies spatially structured emission lines associated with the NLR and the host galaxy.

The technique is particularly well suited to observations of luminous, high-redshift ($z>1$) QSOs obtained with modern IFS facilities such as JWST/NIRSpec (\citealt{Boker2022}). At these redshifts, the host galaxy is often only marginally resolved, or even unresolved, and the detection of extended ionised gas generally relies on an accurate subtraction of the bright nuclear point-spread function (PSF). However, PSF reconstruction is often challenging, especially in the absence of dedicated PSF reference observations or isolated point sources within the field of view, and residual uncertainties can significantly affect the recovered morphology and kinematics of the extended emission (e.g. \citealt{Marshall2024}). The sRMS technique offers a complementary, data-driven approach that exploits the intrinsic spatial information contained in the IFS cube, reducing the reliance on an explicit PSF model. In this paper, we investigate the performance of this method using JWST/NIRSpec observations of two $z\sim6.5$ type~1 QSOs, complemented by realistic simulated data cubes designed to quantify its capabilities and limitations.

 This paper is organised as follows. In Section~\ref{sec:nirspecdata} we describe the NIRSpec IFS observations and the data reduction. In Sect.~\ref{sec:technique} we introduce the sRMS technique and describe its implementation. In Sect.~\ref{sec:ETC} we use realistic NIRSpec simulations to optimise the extraction geometry and assess the performance of the method under different observational configurations. In Sect.~\ref{sec:results} we apply the sRMS technique to the observed $z\sim6.5$ broad-line QSOs and compare the results with conventional spectral analyses. In Sect.~\ref{sec:otherapplications} we discuss broader applications of the method, including the identification of close dual AGNs and the measurement of accurate systemic redshifts in broad-line AGNs observed with other IFS facilities. Finally, our conclusions are summarised in Sect.~\ref{sec:conclusions}.
Throughout, we adopt a flat $\Lambda$CDM cosmology with $H_0=70$~\kms, $\Omega_\Lambda=0.7$, and $\Omega_m=0.3$.

\section{NIRSpec IFS observations and data reduction}\label{sec:nirspecdata}

The sample used in this paper has been assembled from the survey `Galaxy Assembly with NIRSpec IFS' (GA-NIFS\footnote{\url{https://ga-nifs.github.io}}, e.g. \citealt{Lamperti2024, Marconcini2024, RodriguezDelPino2026,Ulivi2026}), as part of programme \#4528 (PI: Kate Isaak). 
Specifically, the type 1 AGNs considered are: \VDES (\citealt{Reed2017}) and \HSC (\citealt{Matsuoka2016}) at $z\sim 6.5$. 
The detailed analysis of the two individual QSOs, including their ionised-gas kinematics, and environment properties, is presented separately in \citet{Prieto-Jimenez2026}. Here, we instead used these observations as representative high-redshift broad-line AGN to demonstrate and assess the sRMS methodology.

These sources were observed in high-resolution IFS mode using the G395H/F290LP grating/filter combination, providing a wavelength coverage of 2.87--5.27~$\mu$m at a resolving power of R~$\sim 2700$  \citep{Jakobsen2022,Boker2022, Rigby2023}, hence at $z\sim6.5$ covering the rest-frame range $\sim 3800-7000~\AA$ and the most prominent optical emission lines. 
Observations were acquired using the NRSIRS2 readout pattern (\citealt{Rauscher2017}). VDES J0224–4711 was observed using a 13-point medium dither pattern with 9 groups per integration, yielding a total exposure time of 2.4 hours. HSC J0859+0022 was observed using a 10-point dither pattern with 13 groups per integration and a total exposure time of 2.5 hours.

We reduced the data using the JWST pipeline, complemented by custom procedures for the correction of instrumental systematics and residual artefacts. A detailed description of the data reduction and cube construction is provided in Appendix~\ref{app:datareduction}. The final cubes were resampled to $0.05\arcsec$ spaxels, providing adequate sampling of the NIRSpec PSF, which has a typical FWHM of $\sim0.18\arcsec$ over the $3$--$5~\mu$m wavelength range \citep[e.g.][]{Jones2026BlackTHUNDER}. This sampling is important for resolving the spatial variations exploited by the sRMS analysis.
{\cpurple It is also worth noting that a proper correction of the `wiggle' features \citep{Perna2023} is important for the sRMS analysis, as detailed in Appendix~\ref{app:datareduction}.}

\section{The sRMS technique}\label{sec:technique}

Inspired by the RMS spectra commonly employed in reverberation mapping (\citealt{Peterson1998a}), we apply a similar statistical formalism in the spatial rather than the temporal domain. Instead of combining spectra obtained over multiple epochs, the sRMS spectrum is computed from a single-epoch IFS data cube using an ensemble of spectra extracted from partially overlapping circular apertures covering the AGN. 

Two parameters define the extraction geometry:
the radius of each extraction aperture, $r_{\rm ap}$, and the radius of the common region, $r_{\rm com}$, that is enclosed by all extraction apertures (and constrain the number of individual spectra).
Both quantities are expressed in units of the Gaussian width ($\sigma_{\rm PSF}$) of the PSF (hence assuming it can be approximated by a 2D Gaussian). Figure~\ref{fig:figure1} (bottom left) illustrates the aperture configuration on the \VDES white light image, that is the image obtained collapsing the cube along the spectral direction, for $r_{\rm ap}=3.5\sigma_{\rm PSF}$ and $r_{\rm com}=1\sigma_{\rm PSF}$ ({\cpurple corresponding to 45 individual spectra}), adopted as the optimal configuration and justified in the next section. 
The centres of the individual extraction apertures are placed on the centres of spaxels satisfying the chosen $r_{\rm com}$ criterion. 
In the white image, {\cpurple the white crosses identify the centres, and the black circles show the corresponding extraction apertures; the filled green circle marks the common area shared by all spectra.} 

For each aperture, an integrated spectrum is extracted adding all spaxels within it, weighting each spaxel only by the fractional area inside the current circular aperture. All spectra are subsequently continuum-subtracted and normalised before computing the sRMS spectrum. This removes variations in the overall nuclear flux level between apertures 
{\cpurple (off-centre apertures lose more of the PSF wings)}, and mitigates wavelength-dependent PSF effects from contributing to the sRMS signal. 
The continuum is modelled independently with two power-laws in the two NIRSpec detectors using emission-line-free wavelength intervals {\cpurple (indicated with orange areas in Fig.~\ref{fig:figure1}, top-right panel)}.  
The spectra are normalised to the median continuum level measured in  
{\cpurple the interval closest to the emission lines of interest. We verified that using the first rather than the second interval does not significantly affect the resulting sRMS spectrum.} 

The sRMS spectrum is defined as

\begin{equation}
{\rm sRMS}(\lambda)=
\left[
\frac{1}{N-1}
\sum_{i=1}^{N}
\left(F_i(\lambda)-\bar{F}(\lambda)\right)^2
\right]^{1/2},
\end{equation}

where $F_i(\lambda)$ is the continuum-subtracted and normalised spectrum extracted from the $i$-th aperture, $\bar{F}(\lambda)$ is the corresponding mean spectrum, and $N$ is the total number of extracted spectra. {\cpurple Because the apertures are partially overlapping, the extracted spectra are not statistically independent; thus, $N-1$ represents the conventional sample-variance normalization rather than the number of independent spatial measurements. As we will see below, the absolute normalization of sRMS is irrelevant for the derived results.}

   \begin{figure*}[!h]
   \centering
    \includegraphics[width=0.89\textwidth]{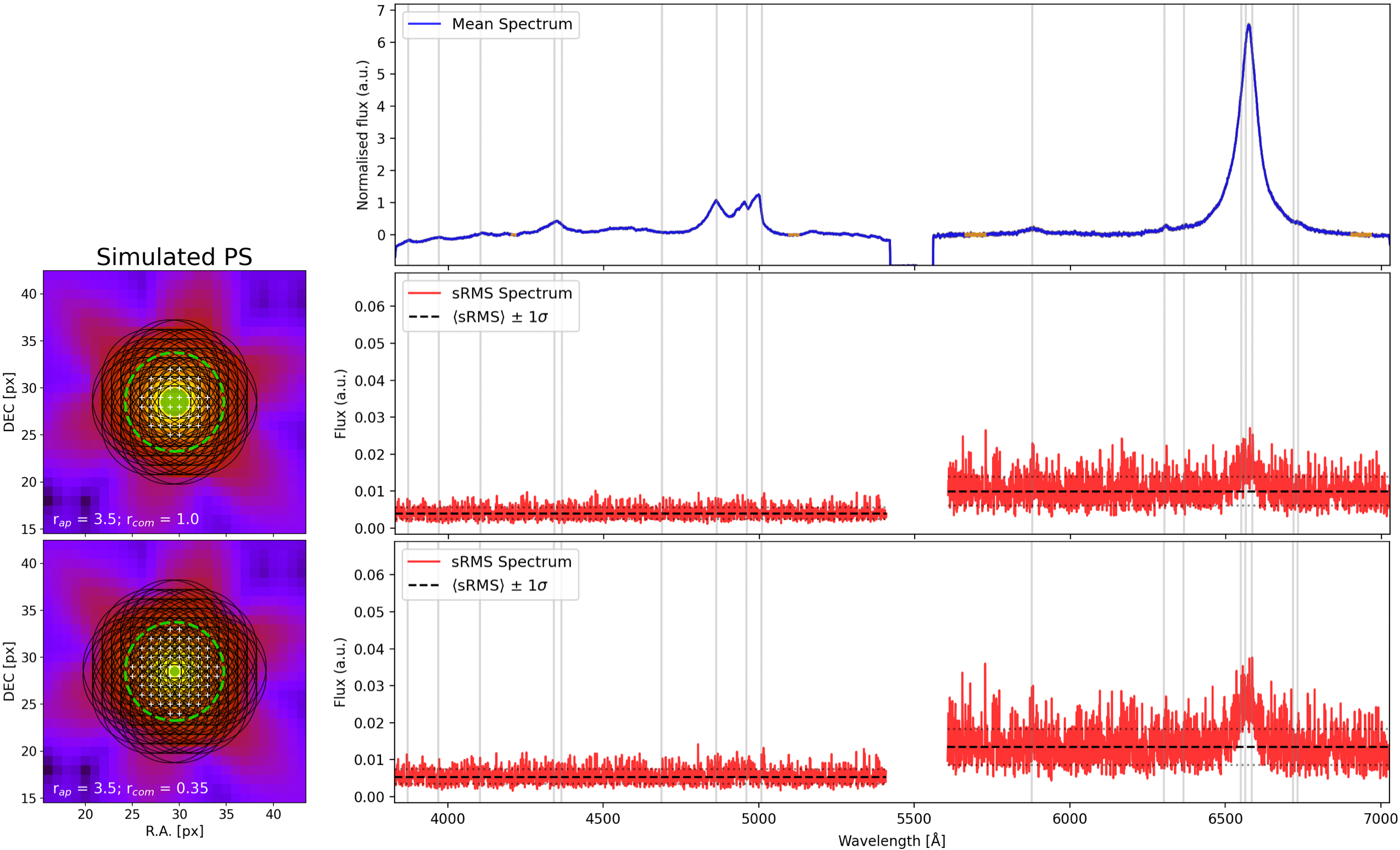}

    \caption{ Application of the sRMS technique to a simulated unresolved type 1 AGN.
    The left panels show white-light images of the simulated NIRSpec IFS cubes for a point-source (PS) target, together with the extraction geometry adopted to compute the sRMS spectrum. The extraction aperture radius is fixed to $r_{\rm ap}=3.5\sigma_{\rm PSF}$, while the common region is set to $r_{\rm com}=1\sigma_{\rm PSF}$ (top) and $r_{\rm com}=0.35\sigma_{\rm PSF}$ (bottom). Black circles indicate the individual extraction apertures, white crosses mark their centres, and the green shaded region corresponds to the common area enclosed by all apertures.
    The right panels show the corresponding mean spectrum (top; displayed for the $r_{\rm com}=1\sigma_{\rm PSF}$ configuration, with negligible differences for the smaller common region) and the resulting sRMS spectra for $r_{\rm com}=1\sigma_{\rm PSF}$ (middle) and $r_{\rm com}=0.35,\sigma_{\rm PSF}$ (bottom). {\cpurple The orange segments in the top panel indicate the regions used for continuum modelling and subtraction.} Since the simulated source is unresolved, the sRMS spectrum efficiently suppresses the line emission. A weak residual broad \ha feature is nevertheless visible, becoming more prominent for the smaller common region owing to the increased spectral differences among the individual extraction apertures.  }
    \label{fig:SIM_PS_fig2}%
    \end{figure*}

The resulting mean and sRMS spectra for \VDES are shown in Fig.~\ref{fig:figure1}. The top-right panel displays the ensemble of individual spectra extracted from the partially overlapping apertures together with their mean spectrum. Although the individual spectra largely overlap, small but significant differences are visible around the narrow emission-line components, most prominently in the \oiii doublet and, to a lesser extent, in the narrow component of \hb (see {\cpurple also} the zoom-in in Fig.~\ref{fig:VDESmeanzoomin}). In contrast, the broad Balmer wings and the \feii emission complexes show no visible variation among the extracted spectra, indicating that these features are dominated by the unresolved nuclear emission and are nearly identical in all apertures. These subtle spectral differences are amplified in the sRMS spectrum shown in the bottom panel of Fig.~\ref{fig:figure1}. The sRMS efficiently suppresses the spatially invariant BLR emission and \feii complexes, while enhancing the spatially varying narrow emission associated with \hb, \oiii, \oi, \ha and \nii. Broad wings associated with the extended ionized outflow remain visible, demonstrating that the sRMS preserves kinematic components that vary across the spatial scales probed. 

{\cpurple It is worth noting that the sRMS spectrum measures the spatial variability of the emission. Consequently, the amplitudes of individual emission-line features in the sRMS spectrum cannot be interpreted as their intrinsic line fluxes. The sRMS is therefore used primarily to identify and constrain the kinematics of spatially varying emission. 
}

As a result, the sRMS provides direct access to the narrow-line emission without requiring an explicit reconstruction and subtraction of the AGN PSF. 
{\cpurple Importantly, unlike conventional PSF subtraction, the sRMS can preserve spatially varying emission in the immediate circumnuclear region. Such emission may be removed together with the unresolved nuclear component when the central spectrum is used to construct and subtract the PSF (Sects.~\ref{sec:psfsub},~\ref{sec:otherapplications}).}

The origin of the spatially varying features and the dependence of the sRMS on the extraction geometry are investigated quantitatively in Sect.~\ref{sec:ETC}. 
In particular, the choice of $r_{\rm ap}$ and $r_{\rm com}$ determines the sensitivity of the sRMS spectrum to compact and extended emission. We therefore optimised these parameters using realistic JWST/NIRSpec simulations before applying the method to the science observations.

   \begin{figure*}
   \centering
    \includegraphics[width=0.89\textwidth]{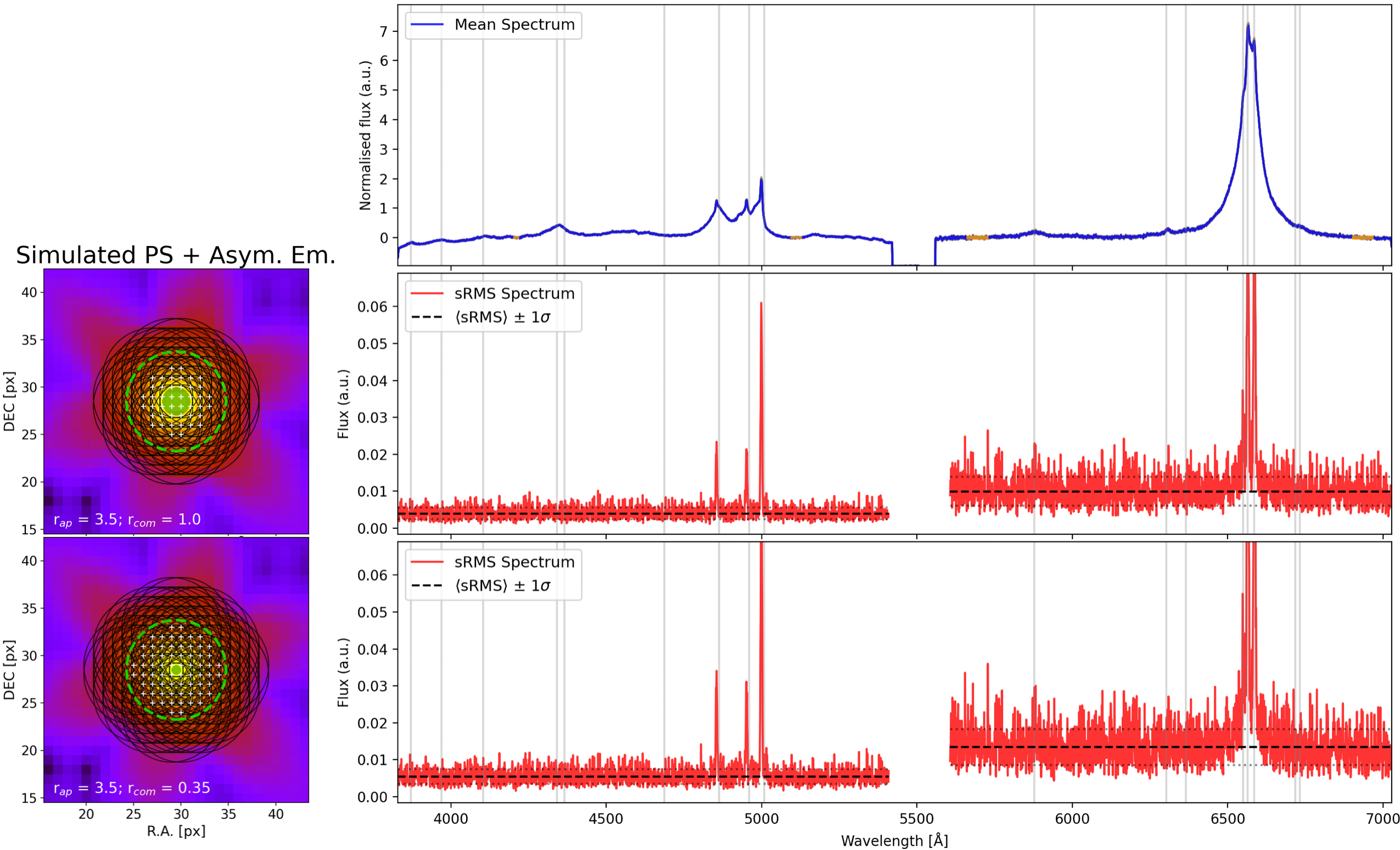}

    \caption{Application of the sRMS technique to a simulated AGN consisting of an unresolved nucleus and spatially offset narrow-line emission.
    The simulated cube contains the same unresolved PS as in Fig.~\ref{fig:SIM_PS_fig2}, together with an extended narrow-line component displaced by $0.04''$ towards the upper-right corner of the field of view. The left panels show the corresponding white-light images and extraction geometry for $r_{\rm ap}=3.5\sigma_{\rm PSF}$, adopting $r_{\rm com}=1\sigma_{\rm PSF}$ (top) and $r_{\rm com}=0.35\sigma_{\rm PSF}$ (bottom). Symbols and colours are the same as in Fig.~\ref{fig:SIM_PS_fig2}.
    The right panels present the mean (top) and the corresponding sRMS spectra obtained with $r_{\rm com}=1,\sigma_{\rm PSF}$ (middle) and $r_{\rm com}=0.35,\sigma_{\rm PSF}$ (bottom). 
    The sRMS spectra clearly recover the spatially varying narrow-line emission (\hb, \oiii, \ha, and \nii), while strongly suppressing unresolved broad-line emission. Reducing the common region slightly increases the significance of the narrow-line features, at the expense of a modest increase in the residual broad-line contribution.   }
    \label{fig:SIM_PSandEXT_fig3}%
    \end{figure*}

\section{Optimisation of the aperture geometry with mock observations}\label{sec:ETC}

To optimise the aperture geometry and quantify the performance of the sRMS technique, we generated realistic JWST/NIRSpec IFS mock observations using version~6.0 of the JWST Exposure Time Calculator (ETC),  
adopting the same observing configuration used for the real observations (Sect.~\ref{sec:nirspecdata}). The ETC-generated simulations were processed into a Python script through the Pandeia engine (\citealt{Pontoppidan2016}). The native $0.1''$ data cubes were resampled to $0.05''$ pixels to match the science observations. Therefore, the resulting simulations reproduce the expected spatial sampling, spectral and angular resolution, and detector noise properties of the science data.

We first generated a point-source-only simulation.
As an input spectrum we adopted the integrated NIRSpec spectrum of VDES~J0224$-$4711. This source exhibits strong overlap between Balmer, forbidden lines, and \feii emission, making it an ideal test case for assessing the capability of the sRMS technique to separate NLR and BLR. 
The observed spectrum was uploaded to the ETC as a point source located at the centre of the NIRSpec field of view. 

To investigate the response of the sRMS technique to extended emission, we generated an additional set of simulations by adding artificial narrow-line emission (\oiii, \hb, \ha\ and \nii) following exponential ($n=1$) Sérsic surface-brightness profiles with specific semi-major and semi-minor axes (with {\cpurple axes ratios} in the range 0.5--1). 
The narrow-line component was assigned fixed kinematic properties (FWHM $=300~$\kms) and fluxes of $6\times 10^{-17}$ \ergs~cm$^{-2}$ for \oiii, \ha and \nii, and $2\times 10^{-17}$ \ergs~cm$^{-2}$ for \hb (hence representing a tiny fraction of total fluxes in the real \VDES spectrum, with $F_{H\beta,~tot}\sim 10^{-15}$~\ergs~cm$^{-2}$). Moreover, this component was displaced with respect to the unresolved nucleus {\cpurple by applying offsets  
smaller than a spatial pixel}. Different {\cpurple offsets and geometrical} configurations were considered, 
to probe spatial variations on scales well below the NIRSpec PSF. Table \ref{tab:sim} summarises their main characteristics, {\cpurple while Fig.~\ref{fig:ETCgeometries} shows two representative configurations used as inputs to the ETC.}

We explored extraction apertures with
$r_{\rm ap}=2$, 3, and 3.5$\sigma_{\rm PSF}$,
combined with
$r_{\rm com}=0.35$, 0.5, 1, 1.5, 2 and 2.5$\sigma_{\rm PSF}$, assuming $\sigma_{\rm PSF} = 0.18\arcsec/2.355 \sim 0.08\arcsec$ (appropriate for the wavelength range investigated in this work; e.g. \citealt{Jones2026BlackTHUNDER}) for all our tests. {\cpurple For the sake of simplicity, we did not consider the variation of the PSF with wavelength.} 
The smallest $r_{\rm com}$ corresponds approximately to a common region of one spaxel ($0.05''$ in diameter), and represents the minimum value considered in this work. The largest aperture considered, $r_{\rm ap}=3.5\sigma_{\rm PSF}$, corresponds to $\sim 5.3$ spaxels and was adopted as a practical upper limit to avoid increasingly incorporating the PSF diffraction spikes visible in the white-light images (e.g. Fig.~\ref{fig:figure1}). 
These two parameters control two competing effects. Increasing $r_{\rm ap}$ increases the total signal collected by each spectrum but also raises the {\cpurple zero-level} of the sRMS spectrum because more detector noise is integrated and because of the contributions from the PSF spikes. 
Conversely, increasing $r_{\rm com}$ makes the extracted spectra progressively more similar, reducing 
the sensitivity to spatially varying narrow-line emission.

Figure~\ref{fig:SIM_PS_fig2} illustrates these effects for simulations containing only the unresolved point source. For both configurations presented in the figure ($r_{\rm com} = 1\sigma_{\rm PSF}$ in central panel, and $r_{\rm com} = 0.35\sigma_{\rm PSF}$ in the bottom panel, for fixed $r_{\rm ap} = 3.5\sigma_{\rm PSF}$), the BLR is efficiently suppressed, although a weak residual broad \ha component remains detectable (at {\cpurple line-peak} S/N~$\sim 2$). {\cpurple For larger $r_{\rm com}$, the residual remains on a floor at S/N~$\sim2$, due to observational noise and residual continuum-modelling differences rather than by spatial variations in the unresolved source.} This residual becomes increasingly prominent for smaller values of $r_{\rm com}$ (up to S/N~$= 2.4$), reflecting the larger differences between the extracted spectra.

The complementary behaviour is illustrated in Fig.~\ref{fig:SIM_PSandEXT_fig3}, where an offset and extended narrow-line component is added to the unresolved source (simulation `{\scriptsize \ding{58}}' in Table~\ref{tab:sim}). The additional contribution is also visible in the integrated spectrum (top panel), showing narrow peaks in \oiii, \ha and \nii emission. In this case the narrow emission lines are clearly detected in the sRMS spectrum (with their kinematic information preserved, see Appendix~\ref{app:ETC}), with the detection significance increasing towards smaller values of $r_{\rm com}$. The improvement is nevertheless modest between $r_{\rm com}=0.35\sigma_{\rm PSF}$ and $1\sigma_{\rm PSF}$; for example, the S/N of \oiii improves only from 36 to 41 in the sRMS spectra shown. {\cpurple Similar emission-line features in the sRMS spectrum are obtained for the other geometrical configurations reported in Table~\ref{tab:sim}}.

{\cpurple Taken together, these controlled simulations demonstrate the basic behaviour of the sRMS technique. In the point-source-only simulations, the sRMS remains largely featureless apart from low-level residuals close to \ha, whereas the addition of spatially varying narrow-line emission produces clear line features in the sRMS spectrum. This demonstrates that the sRMS selectively enhances spectral components with spatial variations across the extraction apertures.}

As a further test, we also considered a simulation in which an extended and symmetric (semi-major and -minor axes equal to 0.15$''$, Sérsic index $n=1$) narrow-line component is centred on the same position as the unresolved nucleus. {\cpurple Because of the absence of a spatial offset, narrow emission lines are barely recovered in the sRMS spectrum, with a much lower significance than in the offset configurations (e.g. \oiii 
${\rm S/N} = 2.4$ for $r_{\rm ap}=3.5\sigma_{\rm PSF}$ and $r_{\rm com}=1\sigma_{\rm PSF}$).} This reduced sensitivity results from the symmetric spatial distribution of the extended component, which produces smaller differences between the spectra extracted at different positions. Such a configuration represents an idealised limiting case rather than a fundamental limitation of the method, since realistic AGN narrow-line regions are expected to exhibit asymmetric spatial distributions, for example as a consequence of obscuration of one side of an ionisation cone, and intrinsic asymmetries in the gas distribution (e.g. \citealt{Fischer2013}).

To determine the optimal extraction geometry we quantified the ratio
$
({\rm S/N})_{\rm NLR}/({\rm S/N})_{\rm BLR},
$
where $({\rm S/N})_{\rm NLR}$ was measured from {\cpurple all} the simulations including both the unresolved nucleus and the offset narrow-line component, while $({\rm S/N})_{\rm BLR}$ was measured from the point-source-only simulations. For the NLR and BLR tracers, we considered \oiii and \ha, respectively. The resulting trends are presented in Appendix~\ref{app:ETC}. 
The ratio $
({\rm S/N})_{\rm NLR}/({\rm S/N})_{\rm BLR}$ reaches its maximum for $r_{\rm ap}=3.5\sigma_{\rm PSF}$ and $r_{\rm com}=1\sigma_{\rm PSF}$. Although broadly comparable values are obtained for smaller common regions, these configurations leave a stronger residual BLR signal. Larger common regions, on the other hand, progressively suppress  
the desired narrow-line emission. We therefore adopt
$r_{\rm ap}=3.5\sigma_{\rm PSF}
{\rm ~and~}
r_{\rm com}=1\sigma_{\rm PSF}
$
as the optimal compromise between efficient suppression of unresolved nuclear emission and high sensitivity to spatially extended narrow-line emission in the sRMS spectra.


   \begin{figure*}[!h]
   \centering
    \includegraphics[width=0.99\textwidth, trim=0mm 2mm 0mm 1mm,clip]{{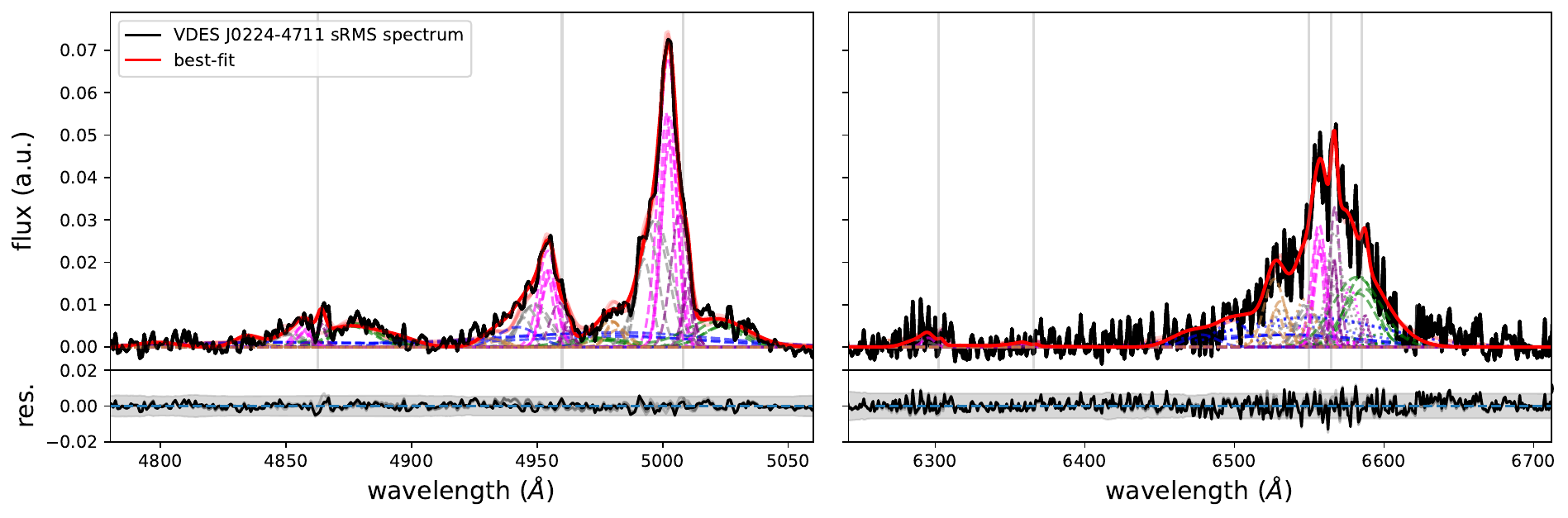}}

    \caption{Kinematic decomposition of the sRMS spectrum of \VDES. The left panels show the \hb--\oiii\ region, while the right panels show the \oi--\ha--\nii\ region. {\cpurple The zero-level was subtracted prior to the kinematic decomposition.} In each panel, {\cpurple three best-fitting models are shown in red, with the six individual Gaussian components displayed in different colours (with dotted curves for \nii and dashed curves for all other lines, reported for each of the three realisations).} The total models are selected among the 100 MC trials, and are shown in light red, while the solution with the lowest $\chi^2$ among them is highlighted in dark red. The lower panels show the corresponding residuals. }
    \label{fig:J0224rmsmodel}%
    \end{figure*}

\section{Application to JWST/NIRSpec observations of broad-line QSOs}\label{sec:results}

The optimal aperture configuration derived above was applied to the reduced JWST/NIRSpec observations described in Sect.~\ref{sec:nirspecdata}. Figure~\ref{fig:figure1} presents the resulting ensemble of extracted spectra together with the corresponding mean and sRMS spectra for \VDES. The corresponding spectra for \HSC are shown in Fig.~\ref{fig:figureA1}. 

We adopt a two-step approach to the spectral decomposition. First, the sRMS spectra are modelled independently of the integrated spectra, providing direct constraints on the kinematics of the spatially varying narrow-line emission. 
These constraints are then used in the decomposition of the integrated spectra, where the spatially unresolved continuum, BLR and other components are modelled. This approach separates the determination of the kinematics of the extended emission from the decomposition of the strongly blended nuclear spectrum, thereby reducing the degeneracy between narrow and broad components. 

We focus here on this aspect of the analysis and on illustrating the potential of the sRMS technique; a detailed analysis of the individual sources and their physical interpretation are presented separately in \citet{Prieto-Jimenez2026}.

\subsection{Kinematic decomposition of the sRMS spectra}\label{sec:sRMSdecomposition}

The first step is to determine the kinematics of the narrow-line emission directly from the sRMS spectra, 
including emission associated with the host galaxy and AGN outflows. We therefore use the sRMS spectrum to identify the number of kinematic components required to describe the extended emission and to determine their velocity centroids and velocity widths. These measurements are subsequently used as constraints when modelling the integrated spectra.

Before fitting the emission lines, we modelled {\cpurple and subtracted the zero-level} using a low-order polynomial, masking the main emission-line regions and the gap between the two detectors. The prominent narrow emission lines were then fitted using the Levenberg--Marquardt least-squares fitting code CAP-MPFIT \citep{Cappellari2017}. We simultaneously modelled the \hb, \oiii, \oi, \ha, and \nii emission-line complexes with Gaussian profiles. All lines associated with a given kinematic component were required to have the same velocity centroid and FWHM, while their amplitudes were allowed to vary between transitions (e.g. \citealt{Perna2020}). The relative fluxes of the \oiii, \oi, and \nii doublet components were fixed to their theoretical ratios \citep{Osterbrock2006}; the sRMS preserves these relative flux ratios in the controlled simulations presented in Appendix~\ref{app:ETC} (and Fig.~\ref{fig:testNLR}), provided that the sRMS zero-level is taken into account {\cpurple (see also Appendix~\ref{app:lineratios})}. The systemic velocity is fixed to that measured from the \cii~158~$\mu$m emission detected with ALMA (\citealt{Wang2024almacii} for \VDES; \citealt{Izumi2018} for \HSC). Rest-frame vacuum wavelengths were adopted throughout.

The number of kinematic components was determined using the Bayesian information criterion (BIC; \citealt{Schwarz1978}), while requiring that each additional component provides a meaningful improvement in the description of the line profiles ($\delta$BIC > 10). 

To assess the robustness of the decomposition and investigate possible degeneracies between individual Gaussian components, we explored the fits using Monte Carlo (MC) realisations. For each source, we generated 100 realisations of the observed sRMS spectrum by perturbing the flux in each spectral channel according to its measured uncertainty, assuming Gaussian errors. For each realisation, the minimisation was performed 500 times, using randomly drawn initial values for the velocity shift ($\Delta v$) and FWHM of each Gaussian component. The initial values were drawn within broad predefined ranges in the $\Delta v$--FWHM parameter space. We deliberately adopted ranges substantially broader than those occupied by the final solutions, in order to minimise the influence of the parameter boundaries and to reduce the possibility that the optimisation converges preferentially to a particular component assignment. The solution with the lowest $\chi^2$ among the 500 initialisations was retained for each realisation. The procedure therefore provides 100 independent best-fitting solutions for each source while reducing the dependence of the results on the choice of initial parameters.

Figure~\ref{fig:J0224rmsmodel} shows the resulting decomposition of the sRMS spectrum of \VDES in the \hb--\oiii and \ha--\nii regions. For clarity, we show only three representative decompositions among the 100 MC solutions. The line profiles are clearly non-Gaussian and require six kinematic components. The narrowest components provide the most direct constraints on the kinematics of the relatively unperturbed spatially varying emission, whereas the broader components trace high-velocity gas associated with the outflow. The component shown in blue is retained in the decomposition for completeness, {\cpurple but is systematically associated with low-amplitudes (with peak S/N $<3$); moreover}, its kinematic parameters are not considered sufficiently robust to be propagated as fixed constraints to the integrated-spectrum modelling.

{\cpurple 
In fact, the MC realisations provide both an estimate of the uncertainties on the individual kinematic components and a test of the stability of the decomposition. Most components show stable solutions across the MC realisations, while one component in each source is found to be substantially less well constrained. This unstable component is shown in blue in Fig.~\ref{fig:J0224rmsmodel}. The detailed comparison of the kinematic solutions in the $\Delta v$--FWHM plane, including the robustness and separation of the individual components, is presented in Sect.~\ref{sec:degeneracy}.
}

For \HSC, the sRMS spectrum similarly reveals spatially varying narrow-line emission with asymmetric profiles (Fig.~\ref{fig:J0859rmsmodel}). The decomposition requires fewer components than for \VDES (i.e. three Gaussians), reflecting the less pronounced complexity of the observed line profiles. The resulting components again provide direct constraints on the kinematics of the extended narrow-line emission, including both the narrow line cores and the broader component associated with the high-velocity gas. As for \VDES, one component is less well constrained (blue Gaussians in the figure) and is therefore not used as an sRMS-derived kinematic prior.

The sRMS decomposition therefore provides an empirically determined set of robust kinematic components that can be carried over to the integrated spectrum. Importantly, the MC analysis also identifies components for which such constraints cannot be reliably obtained. In the following section, we use the robust sRMS solutions as kinematic priors while allowing the integrated spectrum to independently constrain any additional components not securely identified by the sRMS.

\subsection{Constraining the integrated spectra with the sRMS}\label{sec:integratedspecfit}

   \begin{figure*}
   \centering
    \includegraphics[width=0.85\textwidth]{{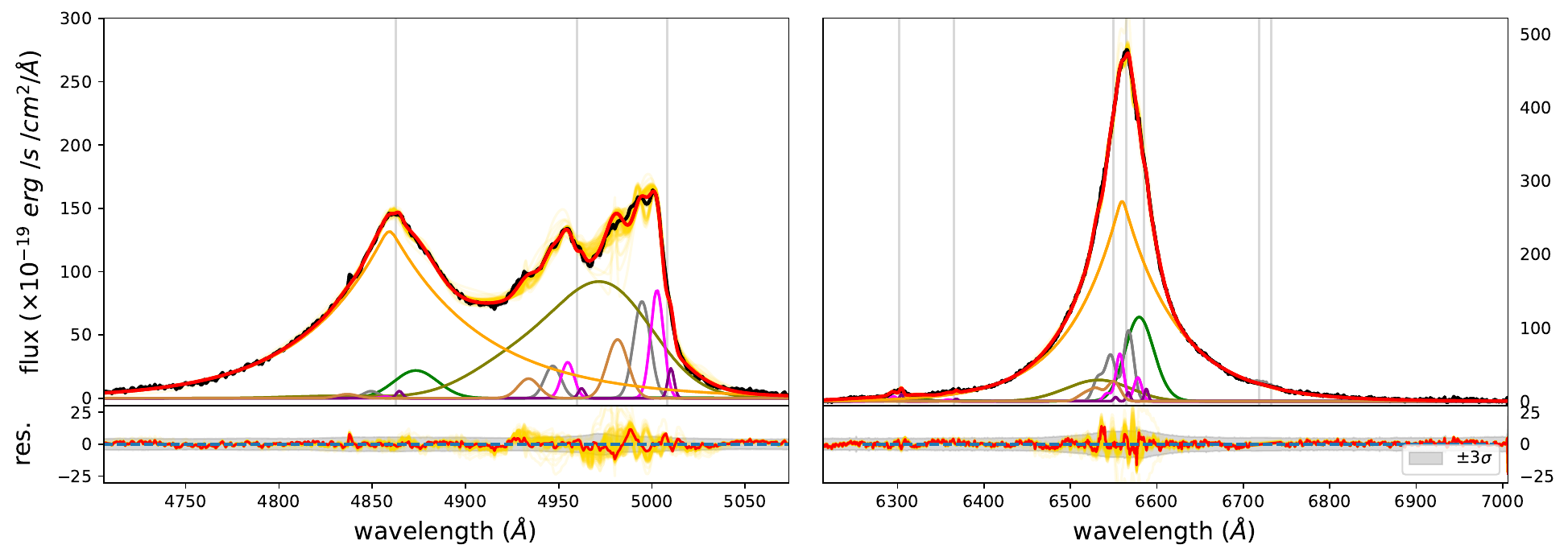}}

    \caption{Kinematic decomposition of the \VDES integrated spectrum. The left panels show the \hb--\oiii\ region, while the right panels show the \ha--\oi--\nii region, after subtracting the continuum and \feii contributions. The total models are selected among the 100 MC realisations and are shown in gold, while the solution with the lowest $\chi^2$ among them is highlighted in red. The lower panels show the corresponding residuals. In each panel, the individual Gaussian components in different colours refer to the realisation associated with the red curve. }
    \label{fig:J0224integmodel}%
    \end{figure*}

The kinematic decomposition of the sRMS spectra provides constraints for the modelling of the spatially integrated AGN spectra. We extracted the integrated spectra using the same circular apertures adopted for the sRMS analysis, with $r_{\rm ap}=3.5\sigma_{\rm PSF}$ centred on the AGN position (i.e. green circles in Figs.~\ref{fig:figure1} and \ref{fig:figureA1}), corresponding to $r_{\rm ap}= 0.27\arcsec$ for the adopted PSF. The integrated spectrum therefore contains the combined contribution of the spatially unresolved nuclear emission and the spatially varying narrow-line emission identified by the sRMS.

Rather than adopting a single kinematic solution from the sRMS decomposition, we propagated the 100 MC solutions described in Sect.~\ref{sec:sRMSdecomposition} to the integrated-spectrum modelling. For each MC solution, the velocity centroid and FWHM of the robust narrow-line components identified in the sRMS were fixed to their corresponding values, while their fluxes were allowed to vary freely. In this way, the full range of kinematic configurations allowed by the sRMS is propagated into the integrated-spectrum decomposition, while avoiding an independent determination of the narrow-line kinematics in the presence of the strongly blended nuclear emission.

The poorly constrained component identified in the sRMS analysis was not used as a kinematic prior. Instead, an additional Gaussian component with free velocity centroid and FWHM was included in the integrated-spectrum model. This provides a flexible way of accounting for emission that contributes to the spatially integrated spectrum but is not robustly constrained by the sRMS. Such a component may arise from spatially compact or approximately symmetric narrow-line emission, which is strongly suppressed by the sRMS, or from residual nuclear emission.  
This treatment therefore avoids imposing unreliable sRMS constraints while allowing the integrated spectrum to determine whether an additional kinematic component is required.

The continuum and \feii emission were modelled following the procedure described in \citet{Prieto-Jimenez2026}. Briefly, the AGN continuum was described using a broken power-law prescription following \citet{Zamora2025}, while the \feii\ emission was modelled using empirical templates from \citet[][in the vicinity of \ha, for \VDES]{Veron2004} and \citet[][in the vicinity of \hb-\oiii for both targets]{Kovacevic2010}. After subtracting the best-fitting continuum and \feii contribution, we focused on the simultaneous modelling of the BLR and narrow emission-line components. The BLR contribution was modelled using broad broken-power-law profiles for both \hb and \ha, while the narrow-line components identified from the sRMS were included with their kinematic parameters fixed to the values obtained from each of the 100 MC solutions. 

Figure~\ref{fig:J0224integmodel} shows the resulting decomposition of the integrated spectrum of \VDES. The 100 fits corresponding to the MC solutions are shown in gold, while the solution providing the lowest $\chi^2$ is highlighted in red together with its individual Gaussian components. The corresponding residuals are shown in the lower panel. 
The analogous decomposition for \HSC is shown in Fig.~\ref{fig:J0859integmodel}.

The constrained models provide a good description of the integrated spectra of each source, despite the fact that only one narrow-line component is allowed to vary freely in its kinematic and flux properties (shown in olive in the figures). The residuals are generally small and are consistent with the integrated spectrum containing emission components that are not efficiently recovered by the sRMS, for instance {\cpurple a collimated outflow along the line of sight.}  
Allowing one component to vary freely provides the flexibility required to account for such emission without compromising the kinematic constraints obtained from the spatially varying components.

This fitting strategy illustrates the main advantage of combining the sRMS and integrated spectral analysis. The sRMS independently identifies the kinematic structure of the spatially varying narrow-line emission, while the integrated spectrum recovers the complete emission-line budget. 
Most of the narrow-line kinematics are therefore determined independently of the strongly blended integrated spectrum, substantially reducing the freedom of the subsequent multi-Gaussian decomposition.

A direct comparison with a conventional unconstrained multi-Gaussian decomposition is presented in Sect.~\ref{sec:degeneracy}. There we show that, although a similar number of Gaussian components can be statistically preferred when fitting the integrated spectrum alone, the unconstrained solutions do not occupy well-defined regions of the $\Delta v$--FWHM parameter space. In contrast, the sRMS-based approach provides physically motivated and well-separated kinematic solutions for the spatially varying components, thereby reducing the degeneracy inherent in the decomposition of the integrated spectrum.

\subsection{Reducing degeneracies in multi-component spectral decomposition}\label{sec:degeneracy}

In this section, we compare the sRMS-constrained decomposition with a conventional approach in which the kinematic parameters of all narrow-line components are allowed to vary freely. We focus on \VDES, the most complex of the two sources and therefore the most compelling case for such a comparison.

   \begin{figure*}
    \text{\tiny \hspace{4.8cm} sRMS constrained \hspace{5cm} Without sRMS constraints}\par\medskip
   \centering
    \includegraphics[width=0.42\textwidth, trim=0mm 3mm 0mm 2mm,clip]{{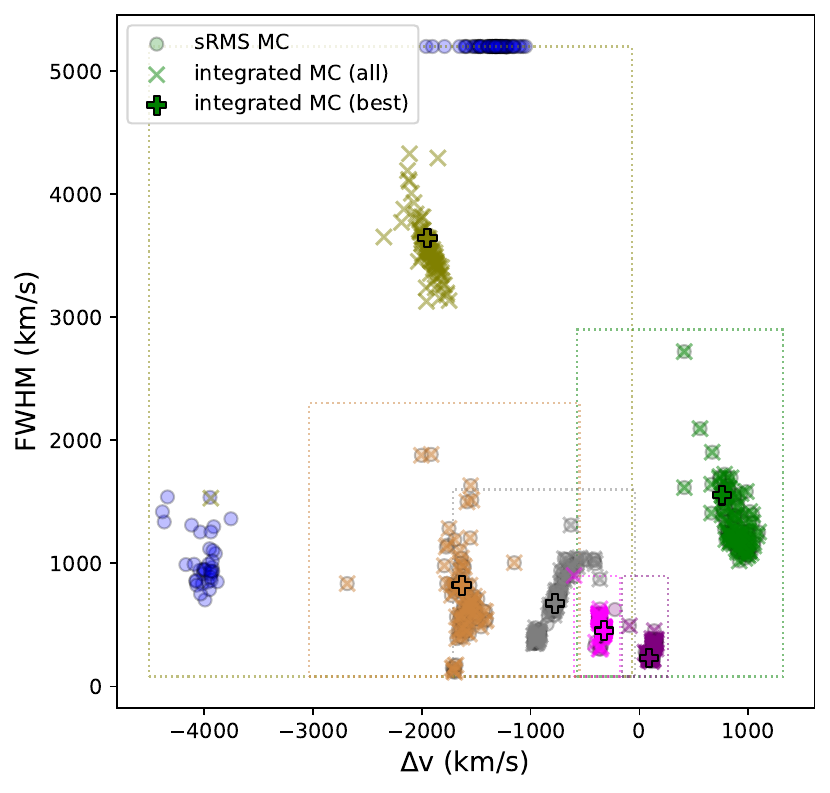}}
    \includegraphics[width=0.42\textwidth, trim=0mm 3mm 0mm 2mm,clip]{{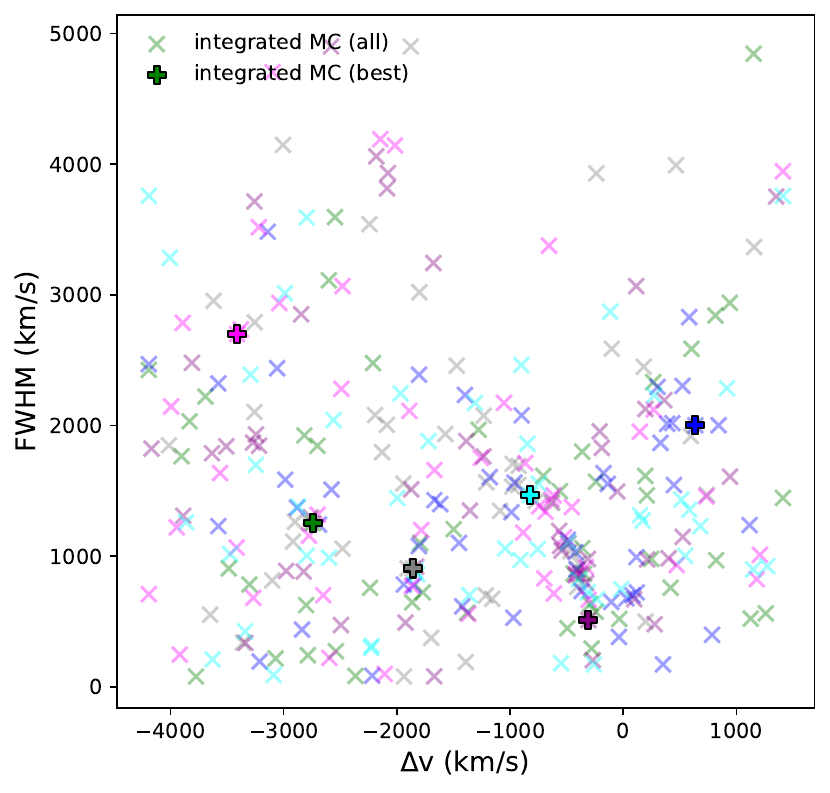}}

    \caption{$\Delta v$ versus FWHM for the Gaussian kinematic components used to model the spatially integrated spectrum of \VDES. Left: fit decomposition obtained from the 100 MC trials with sRMS priors (circles from the fit of sRMS spectra, crosses from the modelling of the integrated spectrum). 
    For the five well-constrained sRMS components, $\Delta v$ and FWHMs are fixed in the integrated-spectrum modelling; consequently, the corresponding crosses overlap the circles in the figure. The sixth component identified in the sRMS analysis (blue circles) is poorly constrained and does not converge to a well-defined region of the plane. {\cpurple Therefore, no blue crosses are presented in the figure; instead, an alternative sixth component is introduced with both $\Delta v$ and FWHM left free, shown by the olive crosses.}  
    The well-separated loci of the individual components illustrate the reduced degeneracy achieved by using the sRMS-derived kinematic constraints. Black plus symbols filled with colours of the specific component refer to the optimal best-fit result shown in Fig.~\ref{fig:J0224integmodel}. Right: corresponding six-component decomposition obtained by fitting the integrated spectrum without sRMS constraints, with all kinematic parameters allowed to vary freely. The resulting solutions are broadly scattered in the plane, illustrating the substantially greater degeneracy of the standard multicomponent decomposition. }
    \label{fig:DVvsFWHM}%
    \end{figure*}

Figure~\ref{fig:DVvsFWHM} (left) shows the distribution of all kinematic components in the $\Delta v$--FWHM parameter space resulting from our new approach. The coloured circles mark the kinematic solutions identified from the sRMS analysis, while the crosses show the corresponding components adopted in the integrated-spectrum decomposition. 
{\cpurple The dotted boxes indicate the parameter ranges used to constrain the individual components during the final Monte Carlo fits, and were introduced to prevent different kinematic components from swapping assignments between realizations (Sect.~\ref{sec:sRMSdecomposition}).}
For the five well-constrained sRMS components, the velocity centroids and FWHMs are fixed in the integrated-spectrum modelling; consequently, the corresponding crosses overlap the circles in the figure. The sixth component identified in the sRMS analysis, shown in blue, is poorly constrained (it is systematically associated with low-amplitude emission, with peak S/N $<3$; see Fig.~\ref{fig:J0224rmsmodel}), and does not converge to a well-defined region of the $\Delta v$--FWHM plane. We therefore do not use its kinematics as a prior in the integrated-spectrum decomposition. Instead, the sixth component is introduced with both $\Delta v$ and FWHM left free, shown by the olive crosses. This component occupies a region of parameter space between the distinct sRMS components and accounts for NLR emission present in the integrated spectrum but not robustly constrained by the sRMS.

For comparison, we fitted the continuum- and \feii-subtracted integrated spectrum without imposing the sRMS kinematic constraints, using models with three to seven freely varying narrow-line Gaussian components and the same BLR prescription adopted in Sect.~\ref{sec:integratedspecfit}. Also in this case, we performed multiple (50) MC realisations (by perturbing the flux and using randomly drawn initial values) to investigate the fit degeneracies. The BIC favours a six-component model, consistent with the total number of components required by the sRMS-based decomposition. The number of Gaussians required by the data is therefore not reduced by the sRMS constraints.

The difference lies instead in the stability and interpretation of the individual components. As shown in the right-hand panel of Fig.~\ref{fig:DVvsFWHM}, the freely fitted components do not cluster around well-defined locations in the $\Delta v$--FWHM plane. Their velocities and widths vary substantially between MC realisations, indicating that several different combinations of Gaussian parameters can provide similarly good descriptions of the integrated line profiles (see also \citealt{Yang2023wfss} for an independent fit of the \hb-\oiii region for the same target). In contrast, the sRMS-based decomposition identifies distinct and well-separated kinematic structures that remain stable across the MC realisations.

This comparison demonstrates that the sRMS approach strongly reduces the kinematic degeneracy. By independently identifying the components associated with spatially varying emission, the sRMS constrains most of the velocity and width parameters before the strongly blended integrated spectrum is modelled. The resulting decomposition therefore provides a more robust basis for associating individual components with distinct spatially extended structures. A detailed physical analysis of the unperturbed and outflow kinematic components is beyond the scope of this work and is presented separately in \citet{Prieto-Jimenez2026}.

\subsection{Validating the sRMS decomposition with PSF-subtracted data}\label{sec:psfsub}

To independently verify that the kinematic components identified in the sRMS spectrum correspond to spatially extended emission, we performed a PSF subtraction of quasar emission in the \VDES NIRSpec/IFU data cube. The analysis was carried out separately for the two NIRSpec detectors, using a PSF model centred on the \hb wavelength for the \hb--\oiii\ spectral region and a PSF model centred on \ha for the corresponding region in detector~2. The results from the two spectral regions were then combined to provide a consistent view of the spatially resolved emission.

The NIRSpec IFS instrumental PSF was generated using the \texttt{STPSF} package (\citealt{Perrin2014}) at the central wavelength of the two Balmer lines. The PSF was initially calculated on an oversampled spatial grid and subsequently resampled to the $0.05\arcsec$ pixel scale of the reduced data cube. It was then rotated to match the orientation of the observations and shifted to the measured position of the nuclear source. The resulting PSF was normalised to its peak value, such that its central pixel had unit amplitude.

   \begin{figure*}
   \centering
    \includegraphics[width=0.82\textwidth, trim=0mm 2mm 0mm 2mm,clip]{{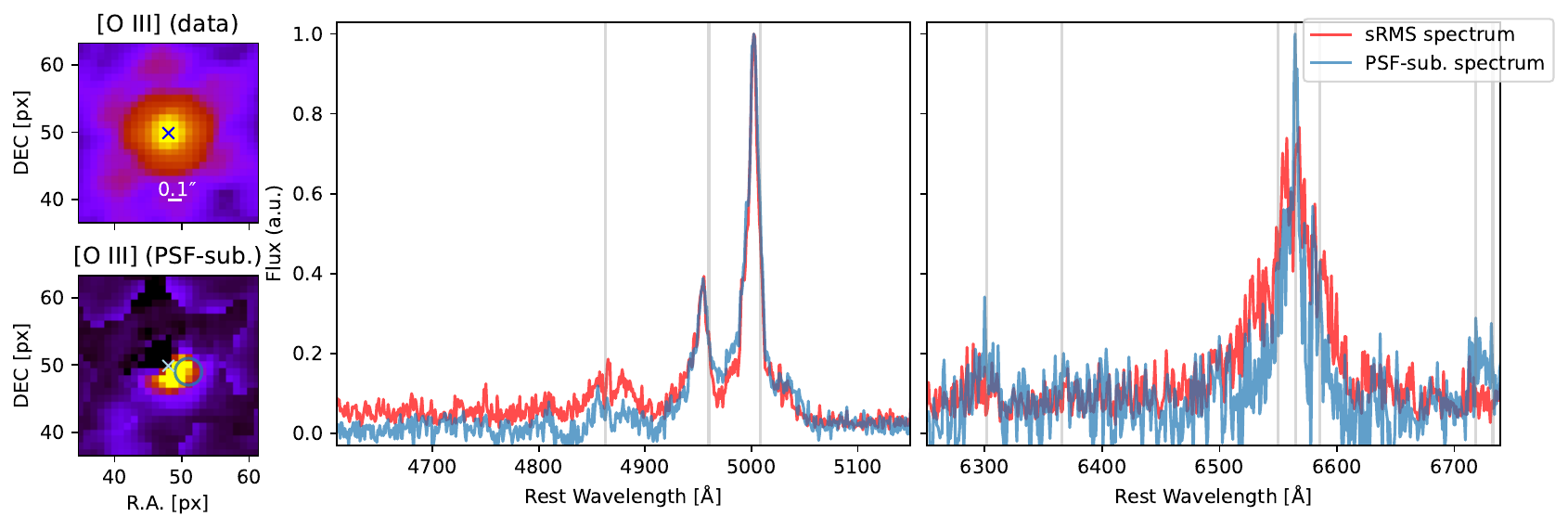}}

    \caption{Comparison between the sRMS spectrum and the PSF-subtracted NIRSpec/IFU data of \VDES. The main panels show the sRMS spectrum extracted from the original data cube (red) and the spectrum extracted from the PSF-subtracted cube (blue) from a region selected to maximise the similarity between the two curves in the \hb-\oiii spectral range. The close agreement demonstrates that the emission-line structures identified in the sRMS are also directly present in the PSF-subtracted data. The top-left inset shows the \oiii flux map from original data cube, while the bottom-left inset shows the \oiii map obtained from the PSF-subtracted cube. In both insets, the position of the nuclear source is marked by an ``$\times$''; the circular aperture in the bottom inset indicates the spatial region used to extract the PSF-subtracted spectrum. }
    \label{fig:VDESPSFsub}%
    \end{figure*}

We extracted the quasar spectrum of the central spaxel, which is dominated by the unresolved AGN component, and used it as a wavelength-dependent template for the PSF. This provides an initial three-dimensional model of the unresolved nuclear emission, obtained by combining the observed nuclear spectrum with the model PSF. However, the resulting model does not reproduce the spectra in the individual spaxels exactly, owing to uncertainties in the PSF model, centring, resampling, and the wavelength dependence of the actual PSF. We therefore introduced a smooth wavelength-dependent correction for each spaxel, parametrised as a low-order polynomial. Its coefficients were determined by minimising the residuals between the observed and model spectra in continuum-dominated spectral regions, masking the main emission lines.
The resulting wavelength-dependent PSF model was then subtracted from the original cube to produce the PSF-subtracted data cube.

Figure~\ref{fig:VDESPSFsub} compares the sRMS spectrum of \VDES with an integrated spectrum extracted from the PSF-subtracted cube. For the latter, we selected a representative spatial position based on visual inspection, where the line structure in the \hb-\oiii spectral range closely resembles that of the sRMS spectrum. The extraction region is smaller than the region sampled by the individual apertures (i.e. $< r_{\rm ap}$), since the sRMS signal is expected to arise from spatial variations within the area covered by the PSF rather than uniformly across the full aperture. The sRMS spectrum is shown in red, while the corresponding spectrum from the PSF-subtracted cube is shown in blue. The two spectra show remarkably similar emission-line structures, providing an independent confirmation that the features identified by the sRMS analysis are present in the spatially resolved data and are not produced by the construction of the sRMS spectrum itself.

The inset panels show the \oiii emission-line images from original and PSF-subtracted data cubes. The \oiii emission extends beyond the unresolved nuclear component, confirming that emission identified through the sRMS analysis is spatially resolved. 
A detailed analysis of the spatial distribution, ionisation properties, and physical origin of these components is beyond the scope of this work and will be presented in \citet{Prieto-Jimenez2026}. Here, we emphasise that, unlike PSF subtraction, which requires an explicit model of the nuclear PSF and its wavelength dependence, the sRMS identifies spatially varying components directly from the observed spectra, without requiring the nuclear contribution to be modelled and subtracted.
{\cpurple Moreover, the sRMS requires no scaling of a nuclear template to each spaxel, and is therefore free from the associated over/under-subtraction degeneracy.}

\subsection{Limitations and applicability of the sRMS technique}

The sRMS spectrum provides information on the spatial variability of the emission, rather than on its mean flux. Consequently, the fluxes of individual emission-line components cannot be directly inferred from the sRMS spectrum and should not be used as quantitative measurements of their intrinsic line luminosities {\cpurple (nevertheless, relative fluxes between nearby emission lines are preserved in the sRMS, see Fig.~\ref{fig:testNLR}).} In particular, a spatially extended component with a relatively axisymmetric surface-brightness with respect to the quasar 
can contribute strongly to the integrated spectrum while producing only a weak sRMS signal. Conversely, a fainter component with a strongly asymmetric spatial distribution can be enhanced in the sRMS, while being  more difficult to identify in the PSF-subtracted data. The primary quantitative information provided by the sRMS is therefore the kinematics of the spatially varying components, which can subsequently be used to constrain their fluxes when modelling the spatially integrated spectrum.

A second limitation concerns the continuum emission. The effectiveness of the sRMS relies on the 
continuum being unresolved and detected at high S/N, so that 
its contribution can be removed consistently from the individual extracted spectra. Spatially varying or noisy continuum emission can instead introduce artificial differences between the normalised spectra. These differences propagate into the sRMS as an elevated zero-level and can produce spectral features even when the corresponding emission-line component is spatially compact. The method is therefore best suited to compact type~1 AGN in which the nuclear continuum is strong and, for instance, dominated by the accretion-disc emission.

This consideration is particularly relevant for recently identified populations such as the so-called ``little red dots'' (LRDs, e.g. \citealt{Matthee2024}), which can exhibit very strong broad emission lines but a comparatively weak optical continuum. In such cases, the noise associated with the continuum modelling and normalisation, as well as non-axisymmetric contributions,  
may dominate the sRMS signal, limiting the ability of the method to distinguish genuinely spatially varying narrow-line emission from continuum-induced variations. 
{\cpurple
We use the case discussed in Appendix~\ref{App:GN9771} to quantify these effects and their dependence on continuum S/N. The applicability of the sRMS technique to LRDs can nevertheless be improved with deeper observations and sources with stronger continuum emission, whose feasibility will be explored in a forthcoming study (Perna et al., in prep.).

}

\section{Additional applications of the sRMS technique}
\label{sec:otherapplications}

    \subsection{Identifying close dual and multiple AGN}

Close dual AGNs are expected to be an important phase of the hierarchical growth of galaxies and their central SMBHs (e.g. \citealt{Volonteri2003,Volonteri2022,Steinborn2016,Rosas-Guevara2019}). Cosmological simulations predict that dual AGN with projected separations below $\sim10$ kpc constitute a non-negligible fraction of the AGN population, reaching a few percent at $z\gtrsim0.5$,  
(e.g. \citealt{PuertoSanchez2024}).
Moreover, theoretical models suggest that AGN activity is preferentially enhanced during the late stages of the merger, when the two SMBHs reach kiloparsec and sub-kiloparsec separations owing to efficient gas inflows toward the nuclear regions (e.g. \citealt{Mayer2007,Capelo2015}). Considerable observational effort has therefore been devoted to identifying close dual AGN in the local and intermediate-redshift Universe (up to $z\sim2-3$; e.g. \citealt{Mannucci2022,Scialpi2024,Scialpi2026}). 
In contrast, only a handful of systems have so far been identified at $z>3$ (e.g. \citealt{Perna2025dual, Perna2026triple,Ubler2024zs7,Ubler2025triple,Zamora2025, Tozzi2026}).

Identifying such systems at $z\gtrsim6$ is particularly challenging. Conventional narrow-line diagnostics, including the widely used BPT (\citealt{Baldwin1981}) diagram, become increasingly difficult to interpret at these redshifts because of the lower metallicities and different ionisation conditions of young galaxies (e.g. \citealt{Feltre2016, Gutkin2016}). Furthermore, in close pairs the narrow-line emission associated with a companion galaxy may be photoionised by the primary AGN, making nebular emission alone an ambiguous tracer of a second accreting SMBH (e.g. \citealt{Marshall2025pearls}). Broad-line emission therefore provides a particularly promising complementary signature. The detection of two spatially distinct BLRs constitutes direct evidence for two actively accreting SMBHs, as already demonstrated for pairs up to $z\sim3$ (e.g. \citealt{Scialpi2026}) and, more recently, in the $z\sim5$ system reported by \citet{Ubler2025triple}. 

The sRMS technique offers a way to search for such systems even when the two nuclei are not spatially resolved. Because the sRMS is sensitive to spatial variations between spectra extracted across the PSF, a BLR associated with a secondary AGN can produce an sRMS signal if its position differs from that of the primary nucleus. This potentially extends the search for dual type~1 AGN to separations substantially smaller than the NIRSpec spatial sampling and angular resolution, down to only a few tens of milliarcseconds, as we demonstrate below.

To assess the sensitivity of the sRMS technique to a close dual type~1 AGN, we performed an additional set of simulations based on the \VDES spectrum. We retained the primary point source and introduced a second, spatially offset point source BLR component in the \ha and \hb lines. The secondary source was assigned a broad-line component in both \ha and \hb, with the same velocity as the primary source and FWHM~$=4000~$\kms, but with lower line fluxes. We adopted $F_{\rm H\beta}=1.5\times10^{-16}$~\ergs~cm$^{-2}$ and $F_{\rm H\alpha}=3F_{\rm H\beta}$, corresponding to one tenth of the broad-line fluxes of \VDES (L$_{\rm H\alpha} = 1.5\times 10^{45}$~\ergs; \citealt{Prieto-Jimenez2026}). The secondary source was also assigned a power-law continuum with $F_\lambda\propto\lambda^{-2}$, normalised according to the expected optical continuum luminosity inferred from the broad \hb luminosity following \citet{Greene2005}. The secondary point source was placed along the spatial $x$-axis at projected separations of {\cpurple 20, 30, 40, and 50~mas (i.e. from 110 to 280 pc at the redshift of the quasar), hence up to 50\% of the NIRSpec IFU native pixel of 100~mas}. These separations are 
substantially below the NIRSpec PSF, and therefore represent configurations in which the two nuclei cannot be spatially resolved directly. Table \ref{tab:sim} summarises the main characteristics of the simulations.

\begin{figure}
   \centering
    \includegraphics[width=0.45\textwidth, trim=-4mm 3mm 0mm 0mm,clip]{{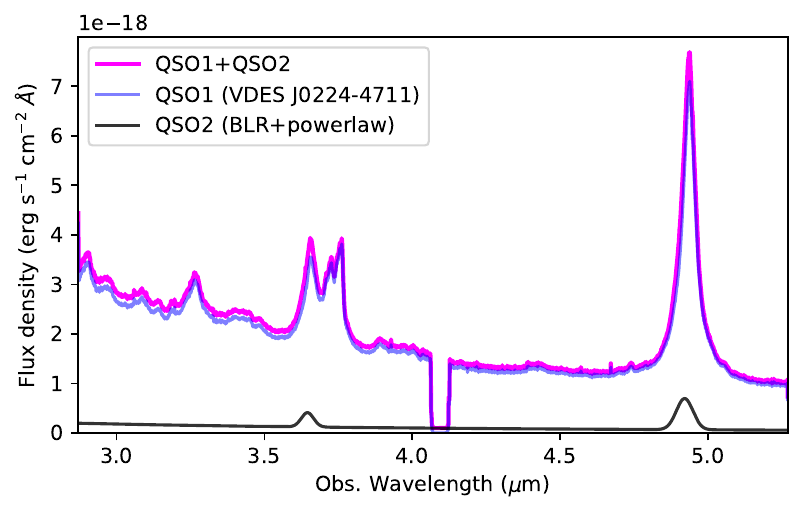}}
    \includegraphics[width=0.45\textwidth, trim=0mm 1mm 0mm 2mm,clip]{{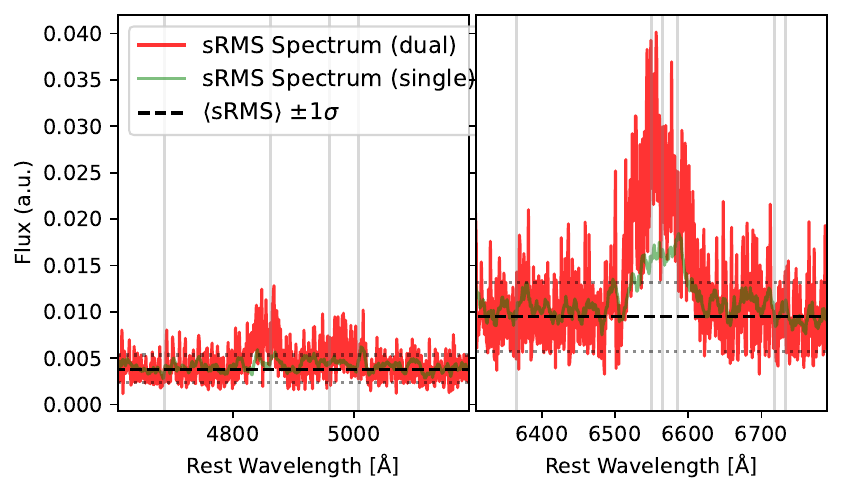}}

    \caption{Simulation of a spatially offset secondary broad-line AGN. The top panel shows the input spectra of the two simulated point sources: the primary source, adopting the observed \VDES spectrum (blue), and the fainter secondary source, consisting of a power-law continuum and broad \hb and \ha emission (black). Their combined spectrum is shown in magenta. The bottom panel shows the resulting sRMS spectrum for a projected separation of $0.04''$. Broad emission is detected at the positions of \hb and \ha, while no comparable signal is produced by the \oiii. {\cpurple A smoothed version of the sRMS for a single AGN (green curve; see Fig.~\ref{fig:SIM_PS_fig2}) is also reported for comparison.}
    }
    \label{fig:dual}%
    \end{figure}

Figure~\ref{fig:dual} illustrates the resulting spectral signature for a separation of $0.04''$. The input spectrum is composed of the \VDES template for the primary source and the fainter broad-line component and continuum of the secondary source (top panel). The corresponding sRMS spectrum shows a clear signal at the positions of \hb and \ha, while no comparable signal is produced by the narrow emission lines such as \oiii (bottom panels). This behaviour is qualitatively different from the simulations of single point-source and spatially varying narrow-line emission simulations presented in Sect.~\ref{sec:ETC}: a spatially offset secondary BLR produces broad Balmer features in the sRMS spectrum without the associated narrow-line emission. 
Therefore, the presence of broad \ha and \hb Balmer emission in the sRMS spectrum provides a distinctive signature that can be used to identify candidate close pairs of broad-line AGN.

The resulting sRMS spectra demonstrate that a spatially offset secondary BLR can be detected even when the two sources are not spatially resolved. For the adopted flux ratio and observing configuration, {\cpurple the \ha component reaches ${\rm S/N}\simeq2.7, 4.6$, 6.0, 7.5, at separations of 0.02, 0.03, 0.04, and 0.05\arcsec\ respectively (compared with ${\rm S/N}\simeq1.8$ for the residual \ha BLR component in the single-AGN simulation, see Sect.~\ref{sec:ETC}).} 
The sRMS signal therefore increases as the spatial separation between the two BLRs becomes larger, as expected from the increasing difference between the spectra extracted at different positions across the PSF.

These simulations illustrate a complementary application of the sRMS technique to searches for close dual type~1 AGN. In contrast to narrow-line diagnostics, which can be affected by the ionisation field of the primary AGN, a spatially varying BLR component directly traces an independently accreting nucleus. Under observing conditions comparable to those adopted here, the simulations indicate that NIRSpec observations can detect a secondary type~1 AGN with broad-line fluxes approximately ten times fainter than those of the primary source at projected separations of order $\sim200$~pc, corresponding to $\sim30$~mas at the redshift of \VDES, for an exposure time of $\sim 2$~hours. This sensitivity is model-dependent and applies to secondary AGN with spectral properties similar to those adopted in the simulations. Nevertheless, it is important to highlight that the sRMS method detects the presence of close secondary AGN in a purely empirical manner, without the need for detailed PSF modelling.

\subsection{Systemic redshifts from UV emission lines in broad-line AGN}

The sRMS technique can also be applied to IFS observations at lower redshift and with instruments other than JWST/NIRSpec. An interesting application is the determination of accurate systemic redshifts for broad-line AGN at cosmic noon, where the bright BLR can dominate the rest-frame UV spectrum and supersede the NLR features that are normally used as systemic-redshift tracers. We illustrate this potential using the QSO J1358+1145 at $z\simeq1.48$, observed both in the Sloan Digital Sky Survey (SDSS) and with the VLT/MUSE instrument as part of the MEGAFLOW survey \citep{Bouche2025}. 

The SDSS and MUSE spectra of J1358+1145 are dominated by broad emission-line profiles, including \civ, \heii, \ciii, and \mgii, making the determination of the systemic redshift from the broad-line peaks intrinsically uncertain. In principle, the narrow \mgii emission provides a more reliable systemic-redshift indicator because \mgii is a low-ionisation transition and is expected to trace gas with relatively small bulk velocities. In practice, however, identifying this component in the integrated spectrum requires disentangling it from the much stronger BLR emission and from the underlying \feii pseudo-continuum. The sRMS provides a direct way to overcome this limitation by suppressing the BLR and enhancing spatially varying narrow emission.

\begin{figure}
   \centering
    \includegraphics[width=0.49\textwidth, trim=0mm 0mm 0mm 0mm,clip]{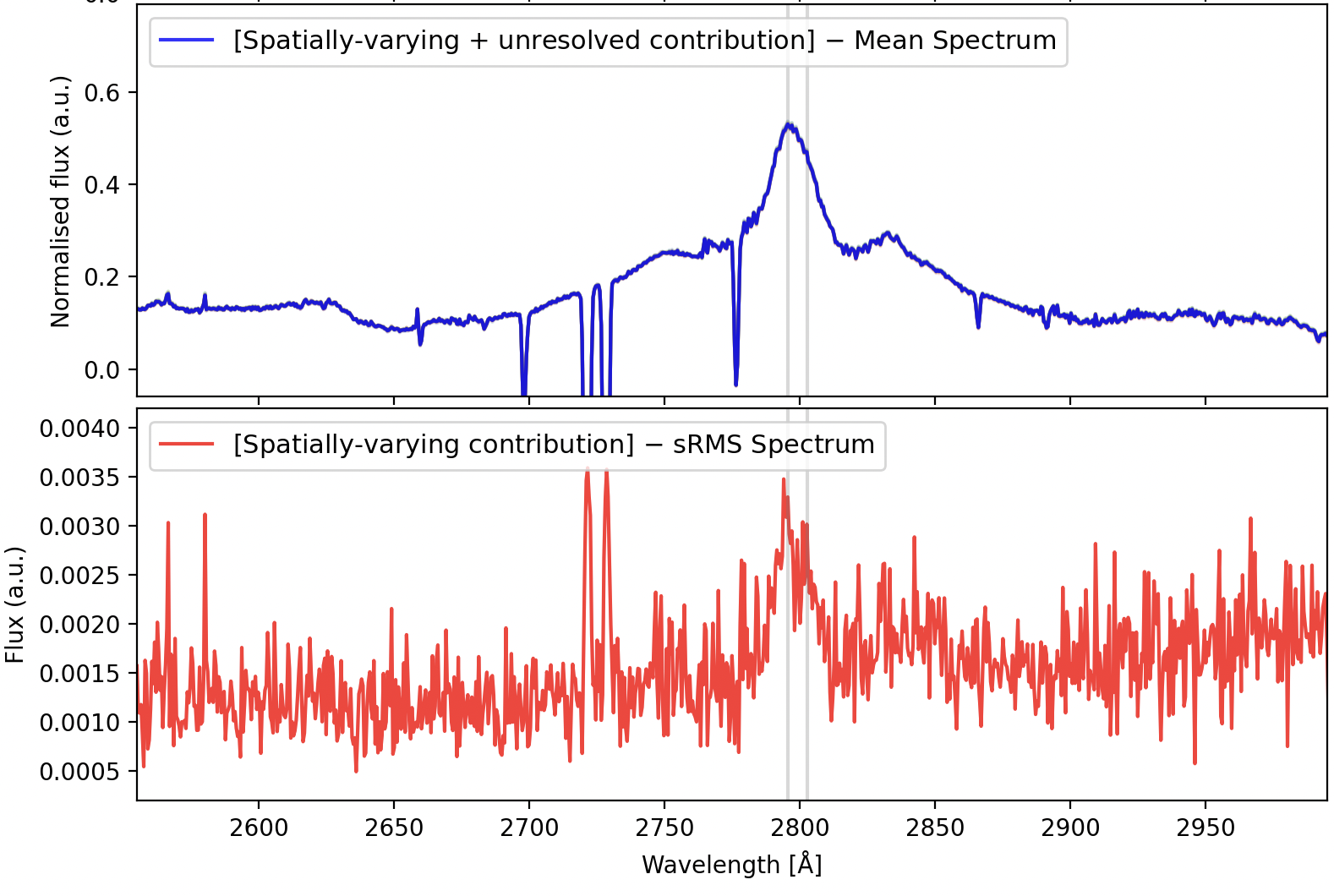}

    \caption{Mean and sRMS spectra of J1358+1145 ($z \sim 1.48$) around the \mgii $\lambda\lambda2796,2803$ doublet, observed with VLT/MUSE. The mean spectrum is dominated by the broad \mgii and \feii emission, while the sRMS spectrum suppresses the spatially unresolved broad components and reveals the narrower \mgii emission components. The sRMS also highlights narrow absorption features (in emission in the sRMS spectrum) at $\sim 2730$~\AA, that are spatially varying across the MUSE field of view.}
    \label{fig:mgii}%
    \end{figure}

Figure~\ref{fig:mgii} shows the mean and sRMS spectra around the \mgii doublet of J1358+1145. The sRMS spectrum reveals the narrow emission components that are largely hidden by the broad profile in the integrated spectrum, providing an independent constraint on the systemic velocity. In addition, the sRMS spectrum highlights narrow absorption features {\cpurple (which appear in emission in the sRMS spectrum)} that are not associated with the quasar. In J1358+1145, the prominent absorption features detected in the sRMS spectrum are therefore spatially varying and are more naturally associated with extended/intervening absorbing structures rather than with a purely nuclear outflow. Their velocity offsets ($\Delta v\sim -8000$~\kms) can consequently be measured relative to the systemic redshift defined by the narrow \mgii emission, $z = 1.4834$. Visual inspection of the MUSE data reveals a galaxy located $\sim3.6\arcsec$ from the quasar whose spectrum shows \mgii absorption features at similar velocities, providing direct evidence that the gas is associated with the circum-galactic medium of a foreground galaxy rather than with a nuclear outflow.

This example illustrates a complementary application of the sRMS technique: by separating spatially varying low-ionisation emission and absorption from the dominant BLR spectrum, it can provide more reliable systemic-redshift constraints for broad-line AGN and help distinguish nuclear spectral features from spatially extended absorbing structures. The method is particularly relevant for IFS surveys of QSOs where conventional systemic-redshift tracers are blended with strong BLR emission.

\section{Conclusions}\label{sec:conclusions}

We have presented the sRMS technique, a method designed to isolate spatially varying emission in IFS observations, and we have applied it to the study of broad-line AGNs. Inspired by the use of RMS spectra in reverberation mapping, the sRMS exploits spatial rather than temporal variations within a single-epoch data cube. By constructing an ensemble of partially overlapping spectra centred on the unresolved nucleus, the method suppresses emission that is spatially invariant across the apertures, while identifying the emission associated with the spatially varying components in the NLR, including outflows and other off-nuclear structures. We applied the method to JWST/NIRSpec IFS observations of two $z\sim6.5$ broad-line QSOs and used realistic simulations to assess its performance.

Our main results can be summarised as follows:

\begin{itemize}

\item The simulations demonstrate that the sRMS can recover narrow-line emission whose spatial offsets with respect to the BLR are substantially ($4-6\times$) smaller than the NIRSpec PSF. 
It can also detect centred  NLR emission, provided its distribution is not perfectly axisymmetric. Therefore, under realistic (asymmetric) NLR configurations, such as those expected for AGN-driven outflows ionisation cones, inhomogeneous dust distribution, and the presence of other offset structures, the sRMS technique can efficiently disentangle the NLR emission from the unresolved BLR component.

\item Applied to the observed QSOs, the sRMS spectra reveal spatially varying narrow emission that is largely blended and hidden by the strong BLR and \feii emission in the integrated spectra. We modelled the sRMS spectra using multiple Gaussian components and explored the robustness of the decomposition through MC realisations. The resulting distributions of velocity offsets and FWHMs remain well separated in the $\Delta v$--FWHM plane for most of the individual components, indicating that the individual kinematic components can be robustly identified despite the complexity of the observed line profiles.

\item The kinematic information derived from the sRMS spectra provides physically motivated constraints for modelling the spatially integrated spectra.  
This approach substantially reduces the degeneracy inherent to multi-Gaussian decompositions, while retaining the flexibility required by the integrated spectra. The comparison with unconstrained fits demonstrates that the sRMS-derived components occupy well-defined regions of the $\Delta v$--FWHM plane; instead, the corresponding components in standard decompositions can vary substantially between different realisations, indicating highly unconstrained kinematic properties in this traditional approach.

\item As an independent validation of the spatial origin of the sRMS features, we performed a PSF subtraction of the \VDES NIRSpec data cube. The residual spectrum extracted from the PSF-subtracted cube reproduces the main kinematic features identified in the sRMS spectrum, demonstrating that the components recovered by the sRMS are also directly present in the spatially resolved emission after removal of the unresolved nuclear PSF. Importantly, the sRMS provides this identification without requiring a detailed modelling and subtraction of the wavelength-dependent PSF, offering a simpler approach for isolating the spatially varying emission.

\item We showed that the technique has applications beyond the identification of off-nuclear and non-axisymmetric narrow-line emission. Simulations indicate that a secondary BLR can produce a detectable sRMS signal even when the two nuclei are separated by  
scales as small as $\sim 200$~pc at $z \sim 6.5$. The resulting broad-line signal has a spectral signature distinct from that of the narrow-line and outflow components recovered in the sRMS, suggesting a promising route to searches for unresolved or barely resolved type~1 dual AGNs at high redshift.

\item We further illustrated the applicability of the technique to other IFS facilities and redshift regimes. In a MUSE observation of the broad-line QSO J1358+1145 at $z\simeq1.48$, the sRMS reveals narrow \mgii emission that is strongly blended with the broad \mgii profile in the integrated spectrum. This provides another example of how the method can help identify low-ionisation narrow emission and constrain the systemic redshift of broad-line AGN. The sRMS also highlights spatially varying absorption features, illustrating its potential for distinguishing spatially extended spectral structures from purely nuclear outflows.
More generally, the technique can be applied to QSOs at $z<1$, provided that the data satisfy the requirements discussed above. In particular, ground-based IFS observations at these redshifts can probe the same rest-frame optical emission-line diagnostics accessible with JWST/NIRSpec IFS for QSOs at $z\sim6.5$.

\end{itemize}

The sRMS technique nevertheless has limitations worth mentioning. First, it does not provide a direct measurement of line fluxes: because the method is based on spatial variations between the extracted spectra, the amplitude of an sRMS feature depends on the spatial distribution of the emission and cannot be straightforwardly converted into an integrated line flux. Its primary quantitative information is therefore the kinematics and spatial variability of the detected components. 
Second, the method requires a sufficiently strong {\cpurple(i.e. detected at S/N $\gtrsim 30$)} and unresolved nuclear continuum. Sources with faint or spatially extended and variable continuum emission can introduce significant variations in the normalised spectra, producing an elevated sRMS baseline and potentially generating apparent line features even when the corresponding emission is spatially unresolved. This limitation is particularly relevant for populations such as the recently identified LRDs, whose spectra can be dominated by strong BLR emission but have relatively weak nuclear continua.

Overall, the sRMS technique provides a simple way of exploiting the spatial information contained in IFS observations to disentangle unresolved and resolved non-axisymmetric emission. Its ability to constrain the kinematics of extended components before fitting the spatially integrated spectrum offers a practical route to reducing degeneracies in complex AGN spectral decompositions. With the increasing availability of high-resolution IFS observations from JWST and current and future ground-based facilities, the sRMS technique can also provide a useful complementary tool for studying AGN-driven outflows, inferring systemic redshifts, and identifying close dual AGN across cosmic time.

{\cpurple Beyond AGN applications, the sRMS technique could also be explored in other IFS contexts where faint spatially varying emission is embedded within a bright unresolved source, such as the detection of binary stars, substellar components, protoplanetary discs, or debris structures (e.g. \citealt{Ruffio2024,Worthen2025NatAs}). The potential of the method in these contexts will be investigated in future studies.}

\begin{acknowledgements}

This work is based on observations made with the NASA/ESA/CSA
James Webb Space Telescope. The data were obtained from the Mikulski
Archive for Space Telescopes at the Space Telescope Science Institute, which
is operated by the Association of Universities for Research in Astronomy, Inc.,
under NASA contract NAS 5-03127 for JWST; and from the European JWST
archive (eJWST) operated by the ESAC Science Data Centre (ESDC) of the
European Space Agency. These observations are associated with the programmes \#4528 (PI: Kate Isaak) and \#5664 (PI: Jorryt Matthee).

MP, SA, LC, BRP, PPG, CPJ, and LU acknowledge support from the research projects 
PID2021-127718NB-I00,
PID2024-159902NA-I00, PID2024-158856NA-I00, PID2025-169195NB-I00, and RYC2023-044853-I of the Spanish Ministry of Science and Innovation/State Agency of Research (MCIN/AEI/10.13039/501100011033) and FSE+.
EB acknowledges support from the INAF Fundamental 
Astrophysics programme 2024.
AJB acknowledges funding from the `FirstGalaxies' Advanced Grant from the European Research Council (ERC) under the European Union’s Horizon 2020 research and innovation program (Grant agreement No. 789056).
MPS acknowledges support under grants RYC2021-033094-I and PID2023-146667NB-I00 funded by MCIN/AEI/10.13039/501100011033 and the European Union NextGenerationEU/PRTR.
H\"U acknowledges support by the Max Planck Society through the Lise Meitner Excellence Program. H\"U acknowledges funding by the European Union (ERC APEX, 101164796). Views and opinions expressed are however those of the authors only and do not necessarily reflect those of the European Union or the European Research Council Executive Agency. Neither the European Union nor the granting authority can be held responsible for them.
EB acknowledges support by INAF through the Bando di Ricerca Fondamentale 2024. 

\end{acknowledgements}

\bibliographystyle{aa}
\bibliography{aanda.bib}

\begin{appendix} 

\section{Mean spectrum of \VDES around \hb--\oiii}

\begin{figure}[!h]
   \centering
    \includegraphics[width=0.43\textwidth]{{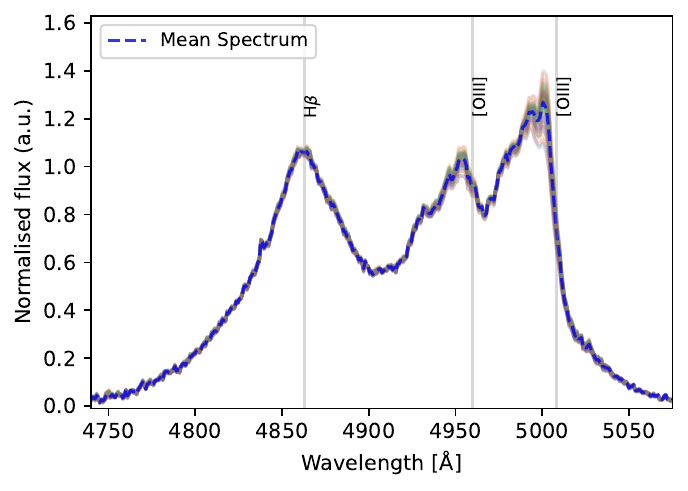}}

    \caption{ Zoom-in of the individual normalised and continuum-subtracted spectra of \VDES around \hb and \oiii. 
    }
    \label{fig:VDESmeanzoomin}%
    \end{figure}

Figure~\ref{fig:VDESmeanzoomin} provides a zoom-in of the individual extracted spectra around the \hb--\oiii complex, also reported in Fig.~\ref{fig:figure1}. This figure highlights the spatial variations around the \oiii peaks while showing no comparable variations across the broad \hb emission. 

\section{Data reduction}\label{app:datareduction}
We downloaded raw data files from the Barbara A. Mikulski Archive for Space Telescopes (MAST) and subsequently processed them with the JWST Science Calibration pipeline (version 2.0.0) under the recommended Calibration Reference Data System (CRDS) context jwst\_1535.pmap. 
In addition to the default pipeline, we applied a set of custom procedures to improve data quality. These include correction of $1/f$ (pink) noise, and masking of residual cosmic rays (including ``snowballs'') and signal from failed open microshutter array shutters. A detailed description of these steps is provided in \citet{Perna2026ring}. 
We adopted the recently developed adaptive trace model (\citealt{Law2026}), which provides a physically motivated, per-spaxel correction for spectral wiggles (see also \citealt{Perna2023}).
We finally generated the cube using the `drizzle' method combined with a standard outlier rejection algorithm, producing a cube with 0.05\arcsec spaxels. This improved resolution is critical for our analysis of the sRMS spectra, as it provides better PSF sampling (the PSF at $3-5~\mu$m is $\sim 0.18''$; see e.g. \citealt{Jones2026BlackTHUNDER}).

{\cpurple It is worth noting that, although wiggles are not expected to significantly affect the spectrum extracted from the aperture centred on the quasar, they can introduce spurious aperture-to-aperture variations in the other spectra. A proper correction of these features is therefore important for the sRMS analysis.
}

\section{Optimal extraction geometry from mock data}\label{app:ETC}

\begin{figure}
   \centering
    \includegraphics[width=0.42\textwidth]{{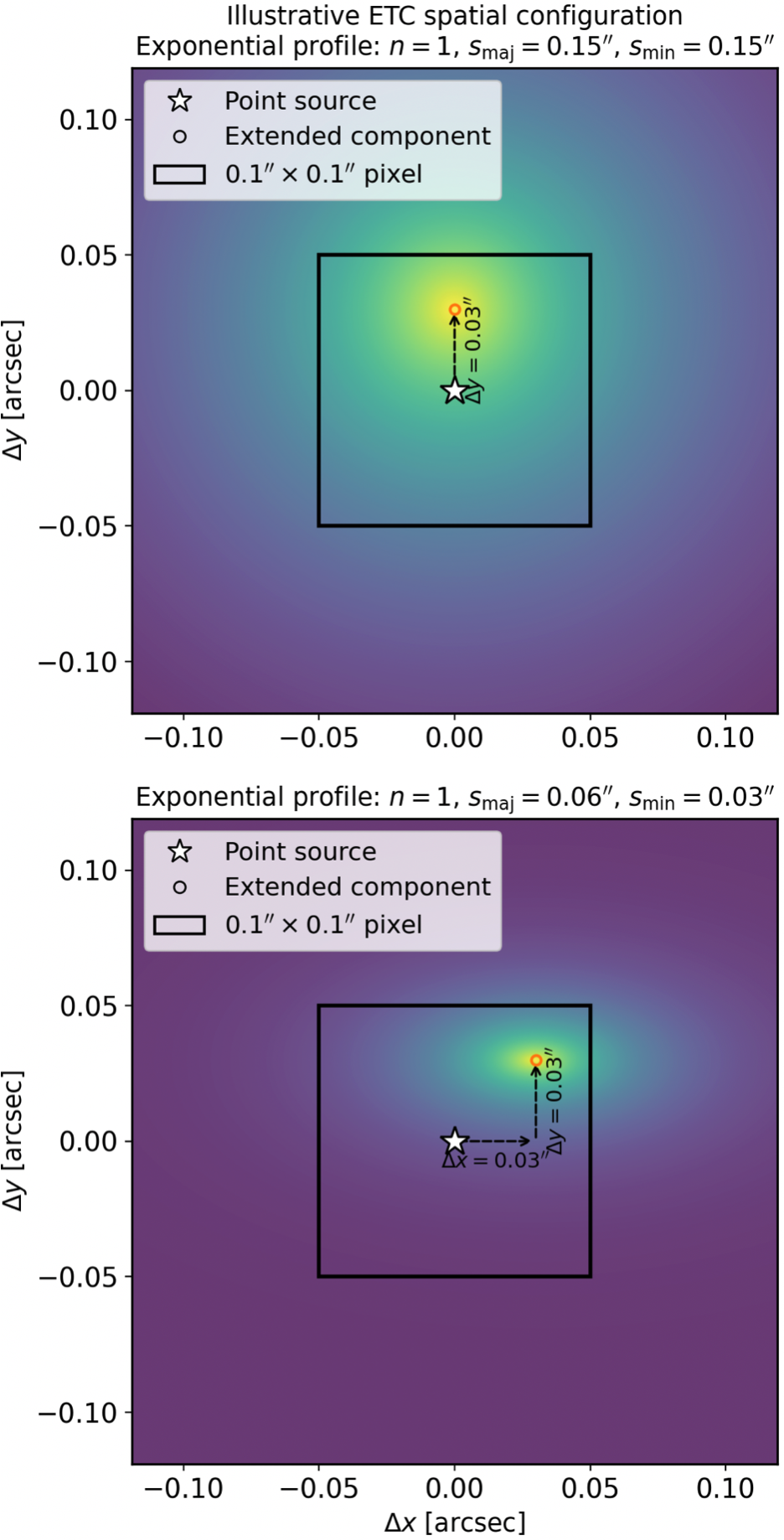}}

    \caption{\cpurple Representative spatial configurations used as inputs to the ETC (with native 0.1$''$ spaxels) and considered for the mock observations. The top panel shows a more extended configuration, with an offset of $+0.03''$ along the $y$-axis. The bottom panel shows a more compact configuration, with an offsets of $+0.03''$ along both the $x$- and $y$-axes. 
    Only the extended component flux distribution is shown, as the central point source (represented by a white star) would overwhelm the former.}
    \label{fig:ETCgeometries}%
    \end{figure}

\begin{figure}
   \centering
    \includegraphics[width=0.46\textwidth]{{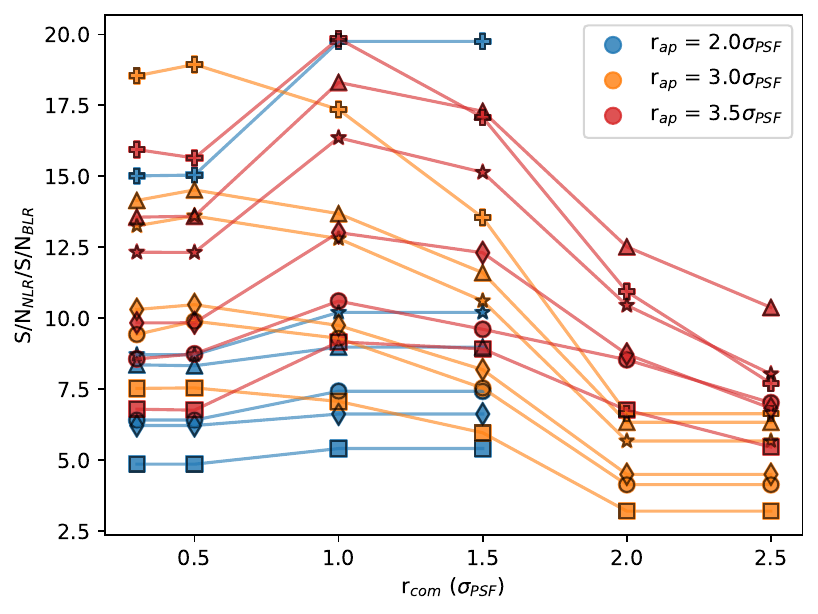}}

    \caption{ $({\rm S/N})_{\rm NLR}/({\rm S/N})_{\rm BLR}$, as a function of the common-region radius, $r_{\rm com}$. Different symbols denote the different spatial configurations of the simulated narrow-line component (Table~\ref{tab:sim}), while colours indicate the adopted aperture radius, $r_{\rm ap}$, as labelled. For each simulation and fixed $r_{\rm ap}$, points corresponding to the different values of $r_{\rm com}$ are connected by lines. The ratio reaches its highest values for $r_{\rm ap}=3.5\sigma_{\rm PSF}$ and $r_{\rm com}=1\sigma_{\rm PSF}$, indicating this configuration as the optimal compromise between the suppression of the unresolved BLR and the recovery of spatially varying narrow-line emission. }
    \label{fig:figureAopt}%
    \end{figure}

We further assess the dependence of the sRMS performance on the extraction geometry using the suite of mock observations described in Sect.~\ref{sec:ETC}.
{\cpurple Figure~\ref{fig:ETCgeometries} illustrates two representative configurations used as inputs to the ETC  
(simulations `{\scriptsize \ding{108}}' and `{\scriptsize \ding{58}}' in Table~\ref{tab:sim}, respectively).}
Figure~\ref{fig:figureAopt} shows the ratio of the signal-to-noise ratios of the narrow-line and broad-line components, $
({\rm S/N})_{\rm NLR}/({\rm S/N})_{\rm BLR}
$, for the simulated spatial offsets and for the different combinations of $r_{\rm ap}$ and $r_{\rm com}$. Differences in S/N ratio values for specific simulations reflect the dependence of the sRMS on the adopted spatial configurations (Table~\ref{tab:sim}).

The results show a broadly consistent preference for an aperture radius of $r_{\rm ap}=3.5\sigma_{\rm PSF}$ combined with a common region of $r_{\rm com}=1\sigma_{\rm PSF}$. Smaller common regions provide a comparable or slightly higher sensitivity to the spatially varying narrow-line emission, but also produce larger variations between the different realisations and leave stronger residuals from the unresolved broad-line emission. Conversely, larger common regions progressively suppress the spatially varying narrow-line signal. The adopted configuration therefore represents a robust compromise between suppression of unresolved nuclear emission and sensitivity to spatially extended emission.
%

%

   \begin{figure}
   \centering
    \includegraphics[width=0.46\textwidth]{{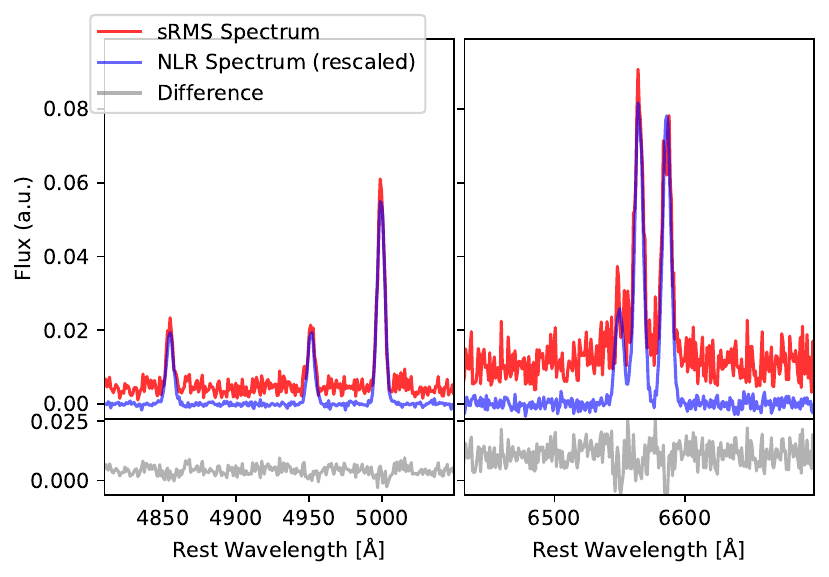}}

    \caption{ Comparison between the sRMS spectrum obtained with the adopted extraction geometry ($r_{\rm ap}=3.5\sigma_{\rm PSF}$, $r_{\rm com}=1\sigma_{\rm PSF}$), and the spectrum of the input spatially extended narrow-line component. The left and right panels show the \hb-\oiii and \ha-\nii regions, respectively. The emission lines recovered in the sRMS spectrum have consistent velocity centroids and widths with those of the input extended component, indicating that the sRMS procedure does not significantly modify the line kinematics. The relative strengths of the \oiii doublet and the \oiii/\hb ratio are also preserved, while larger deviations are observed for the \ha and \nii lines because of their lower S/N. The absence of a broad \ha residual in the sRMS spectrum, also shown in the bottom panels showing the difference between the spectra, further demonstrates the efficient suppression of the unresolved BLR with the adopted extraction geometry. }
    \label{fig:testNLR}%
    \end{figure}

\begin{table*}[!h]
\centering
\caption{Configurations for the second component added to mock data presented in Sects.~\ref{sec:ETC} and \ref{sec:otherapplications}.} 
\label{tab:sim}
\begin{tabular}{l cc ccc ccccc}
\hline
 Sim & $\delta x$ & $\delta y$ & $smin$ & $smaj$ & $n$ & F(\hb) & F(\oiii) & F(\ha) & F(\nii) & $\lambda$F(5100\AA)\\
 & \arcsec & \arcsec & \arcsec & \arcsec & & \multicolumn{5}{c}{($10^{-17}$~\ergs cm$^2$)} \\
\hline
 (1: {\scriptsize \ding{108}}) & 0.03 & 0.03 & 0.15 & 0.15 & 1 & 2 & 6 & 6 & 6 & 0\\
 (2: {\scriptsize \ding{110}}) & 0.00 & 0.03 & 0.15 & 0.15 & 1 & 2 & 6 & 6 & 6 & 0\\
 (3: {\scriptsize \ding{116}}) & --0.03 & --0.03 & 0.15 & 0.15 & 1 & 2 & 6 & 6 & 6 & 0\\
 (4: {\scriptsize \ding{169}}) & 0.00 & --0.03 & 0.15 & 0.15 & 1 & 2 & 6 & 6 & 6 & 0\\
 (5: {\scriptsize \ding{72}}) & 0.00 & --0.03 & 0.10 & 0.15 & 1 & 2 & 6 & 6 & 6 & 0\\
 (6: {\scriptsize \ding{58}}) & +0.03 & +0.03 & 0.03 & 0.06 & 1 & 2 & 6 & 6 & 6 & 0\\
\hline
(7) & +0.03 & +0.00 & -- & -- & -- & 15 & -- & 45 & -- & 420\\
(8) & +0.04 & +0.00 & -- & -- & -- & 15 & -- & 45 & -- & 420\\
(9) & +0.05 & +0.00 & -- & -- & -- & 15 & -- & 45 & -- & 420\\

\hline
\end{tabular}

\tablefoot{Simulations 1--6  used to optimise $r_{\rm ap}$ and $r_{\rm com}$ parameters (Sect.~\ref{sec:ETC}), and associated with the symbols shown in Fig.~\ref{fig:figureAopt}; the remaining simulations correspond to ones used to investigate the detectability of close dual type~1 AGN (Sect.~\ref{sec:otherapplications}). The columns $\delta x$ and $\delta y$ give the spatial offset of the secondary component from the primary source, while $smaj$, $smin$, and $n$ specify the semi-major axis, semi-minor axis, and Sérsic index of the extended narrow-line component, respectively. The latter parameters are not applicable to the 7--9 simulations, where the secondary component is instead modelled as an unresolved broad-line AGN with \ha and \hb BLR emission and a continuum.}
\end{table*}

As an additional test, Fig.~\ref{fig:testNLR} compares the sRMS spectrum obtained from a simulated data cube with a point source (associated with the \VDES integrated spectrum) and an extended narrow-line emission component (simulation `{\scriptsize \ding{58}}' in Table~\ref{tab:sim}), together with the spectrum extracted from a second simulation containing only the extended component. The comparison is shown for the \hb-\oiii and \ha-\nii regions for the realisation in which the extended component is displaced by $0.04''$ from the central point source, and for the adopted configuration ($r_{\rm ap}=3.5\sigma_{\rm PSF}$ and $r_{\rm com}=1\sigma_{\rm PSF}$). The integrated spectrum of the extended component has been rescaled to facilitate the comparison. The emission lines recovered in the sRMS spectrum have the same velocity centroids and widths as those in the input extended component, demonstrating that the sRMS procedure does not introduce significant biases in the measured line kinematics. The relative fluxes of the \oiii doublet and the \oiii/\hb ratio are also preserved, provided that the sRMS zero-level is taken into account (Appendix~\ref{app:lineratios}). The \nii doublet ratio and the \nii/\ha ratio show larger deviations, however, reflecting the lower S/N of these weaker features in the sRMS spectrum. This controlled experiment demonstrates that the sRMS, in addition to the kinematic information, preserves the relative properties of spatially varying emission, although the reliability of line ratios derived from sRMS spectra in real observations will depend on the spatial distribution, noise properties, and relative strength of the individual emission components. {\cpurple We note also that, since the spectral regions in detectors one and two are treated independently, the ratios between lines the two detectors (e.g. \ha/\hb and \ha/\oiii) are not conserved. }

Finally, the sRMS spectrum shows no significant residual contribution from the broad \ha component in the region where the extended narrow-line \ha and \nii emission is recovered. This confirms that $r_{\rm ap}=3.5\sigma_{\rm PSF}$ and $r_{\rm com}=1\sigma_{\rm PSF}$ provide an effective suppression of the spatially unresolved BLR while retaining sensitivity to spatially varying narrow-line emission.

\section{Line fitting  corrections due to the zero-level subtraction in the sRMS spectrum}\label{app:lineratios}

{\cpurple

Figure \ref{fig:testNLR} demonstrates that, for a single kinematic component, the line kinematics and relative line properties can be recovered reliably from the sRMS spectrum. In this controlled case, the line profiles are well separated from the sRMS zero-level, allowing their properties to be measured directly, for instance with simple Gaussian fits masking the wavelength channels dominated by the zero-level contribution. The situation is more complex for the QSO spectra, where some kinematic components can contribute to the sRMS spectrum at levels comparable to the zero-level, making such masking liable to exclude genuine line emission. 
We adopt a two-step approach by modelling and subtracting the sRMS zero-level before fitting the line profiles. This simplifies the identification of multiple kinematic components, but requires small corrections to the theoretical doublet ratios and measured FWHMs.
In fact, the removal of the zero-level in the sRMS spectrum required for the line fitting  may introduce small modifications in the line profiles. In particular,  this subtraction leads to (slightly) different line flux ratios  and widths (i.e. FWHM) with respect to the intrinsic ones, especially if the contribution of the lines to the sRMS spectrum is small with respect to  its zero-level.

Consider a doublet with intrinsic amplitudes $A$ and $a$, with a theoretical ratio $A/a=3$, and a local zero-level $c$. After subtracting the zero-level, the measured amplitudes are $A'=A-c$ and $a'=a-c$, such that the theoretical ratio is expressed as
$(A'+c)/(a'+c)=3$. %
We therefore account for the zero-level subtraction when imposing the theoretical ratios of the \oiii, \oi, and \nii doublets in the sRMS fits. 

Similarly, the FWHM measured after zero-level subtraction is underestimated by a factor
\begin{equation*}
\left[\frac{\ln\left(2A'/(A'-c)\right)}{\ln 2}\right]^{1/2}.
\end{equation*}

We account for this effect by using the \oiii$\lambda5008$ line amplitude to correct the FWHMs inferred from the fits of the sRMS before propagating these results to the modelling of the integrated spectra of the quasars \VDES and \HSC. 
The correction in FWHM corresponds to a factor of $\sim 1.15-1.30$ for \HSC and $\sim 1.06-1.20$ for \VDES; therefore, these corrections are smaller than the uncertainties we derive from the MC trials (see Fig.~\ref{fig:DVvsFWHM}, left). 
}

\section{\HSC mean and sRMS spectra}

\begin{figure*}[!hp]
   \centering
   \includegraphics[width=0.25\textwidth, trim=0mm -60mm 0mm 0mm,clip]{{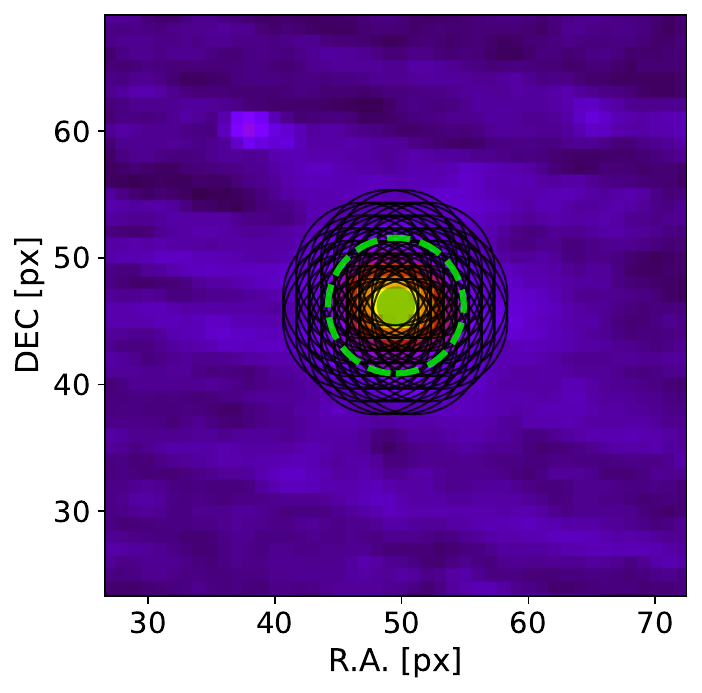}}
    \includegraphics[width=0.7\textwidth]{{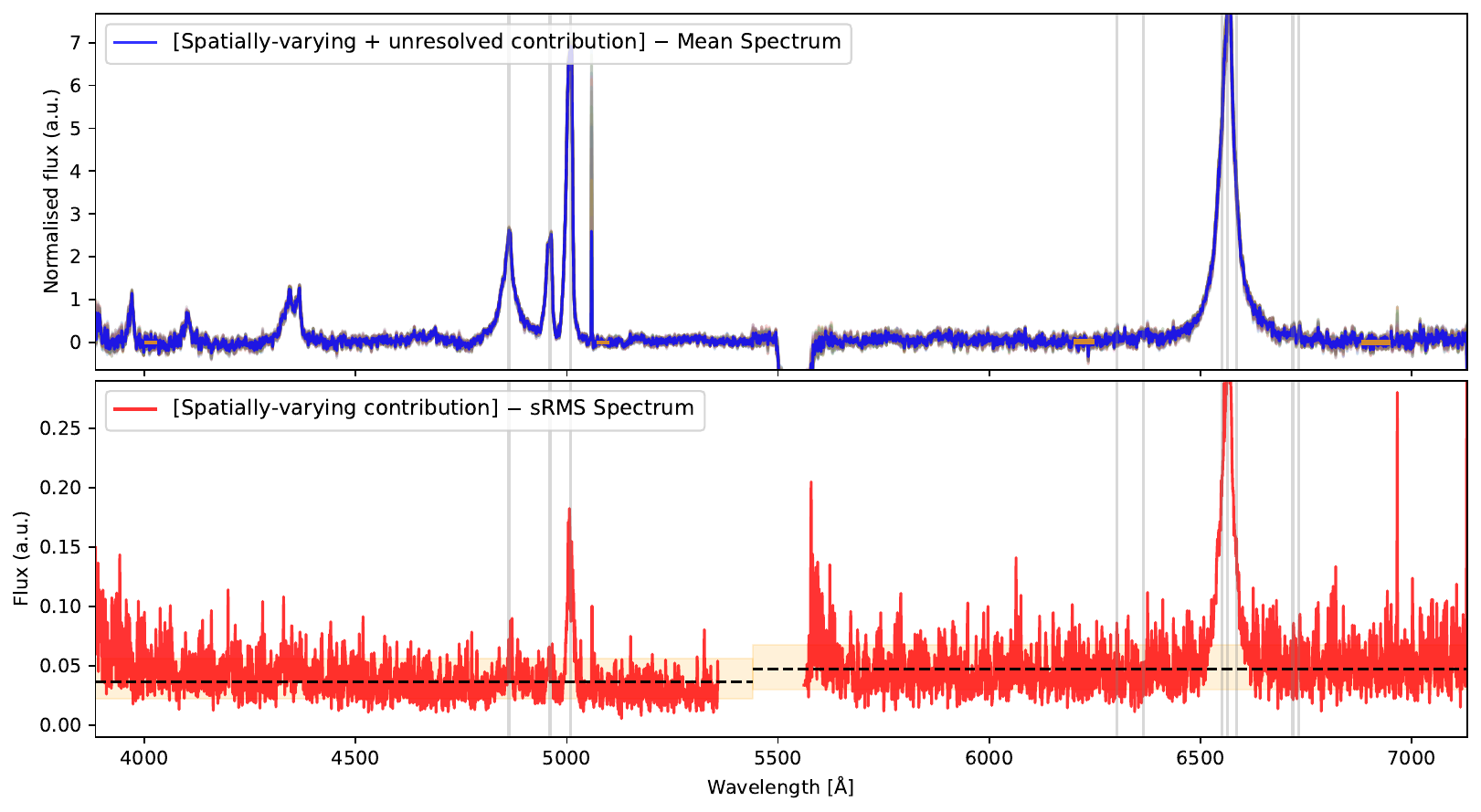}}

    \caption{  \HSC white image, mean and sRMS spectra from JWST/NIRSpec IFS data. See Fig.~\ref{fig:figure1} for detailed description. Also for this target, the sRMS spectrum highlights the spatially varying emission associated with the narrow emission lines \hb, \oiii, \ha, and \nii, while largely suppressing the spatially invariant emission from the unresolved AGN nucleus. }
    \label{fig:figureA1}%
    \end{figure*}

Figure~\ref{fig:figureA1} presents the mean and sRMS spectra obtained for \HSC using the optimal extraction geometry derived from the mock observations. As for \VDES, the sRMS spectrum strongly suppresses the spatially unresolved continuum and BLR emission, while enhancing emission associated with spatially varying narrow-line components.

We first model the sRMS spectrum following the procedure described in Sect.~\ref{sec:sRMSdecomposition}. The decomposition of the main emission-line complexes provides the kinematic components used to constrain the subsequent modelling of the integrated spectrum. The MC realisations described in Sect.~\ref{sec:sRMSdecomposition} are used to assess the robustness of these components and to propagate the range of allowed $\Delta V$ and FWHMs into the integrated-spectrum fit.

The resulting decomposition of the integrated spectrum is shown in Fig.~\ref{fig:J0859integmodel}. The narrow-line components identified from the sRMS analysis are included with their kinematic parameters fixed to each of the MC solutions, while their fluxes remain free. As for \VDES, the integrated spectrum requires additional kinematic components that are not detected in the sRMS spectrum. We therefore include two additional Gaussian components for all the main optical line species. The resulting model provides a good description of the observed emission-line profiles while preserving the kinematic constraints imposed by the spatially varying emission identified in the sRMS spectrum.

%

   \begin{figure*}
   \centering
    \includegraphics[width=0.85\textwidth]{{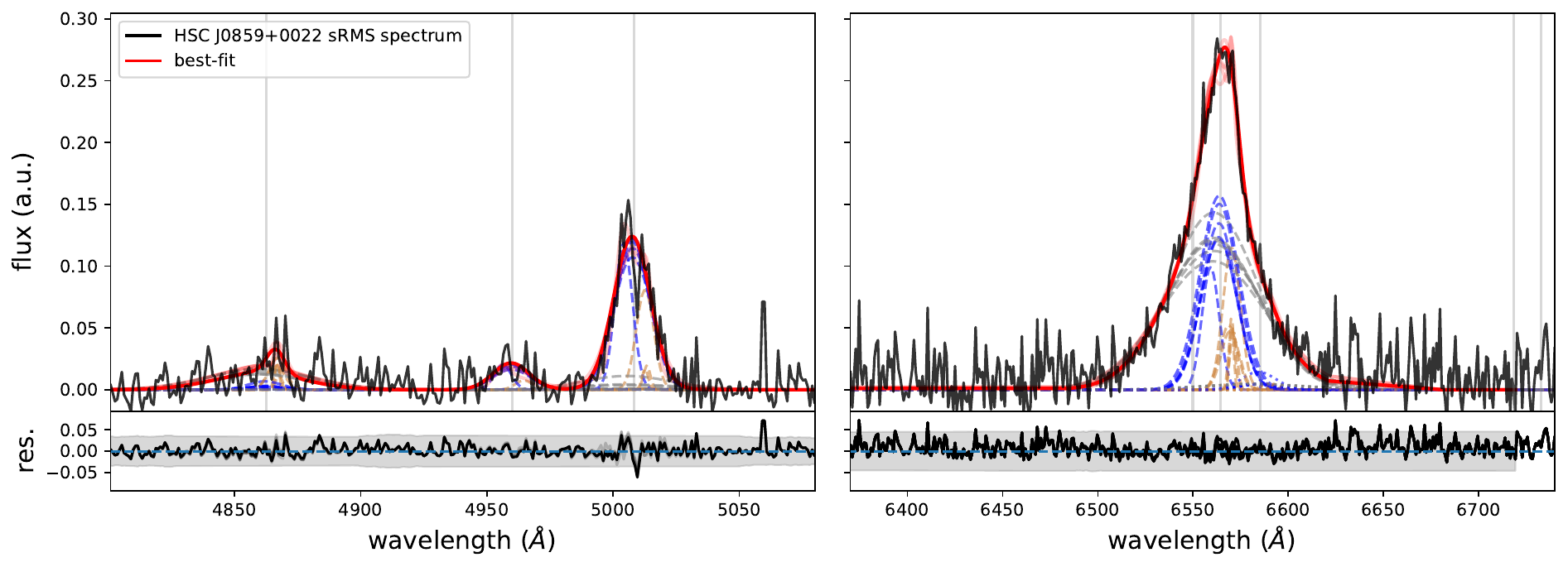}}

    \caption{Kinematic decomposition of the \HSC sRMS spectrum. The left panels show the \hb--\oiii\ region, while the right panels show the \ha--\oi--\nii\ region. {\cpurple The zero-level was subtracted prior to the kinematic decomposition.} In each panel, three best-fitting models are shown in red, with the individual Gaussian components displayed in different colours (with dotted curves for \nii and dashed curves for all other lines). See Fig.~\ref{fig:J0224rmsmodel} for further details. 
    The lower panels show 
    the residual of the best-fitting solution. 
    }
    \label{fig:J0859rmsmodel}%
    \end{figure*}
%

   \begin{figure*}
   \centering
    \includegraphics[width=0.85\textwidth]{{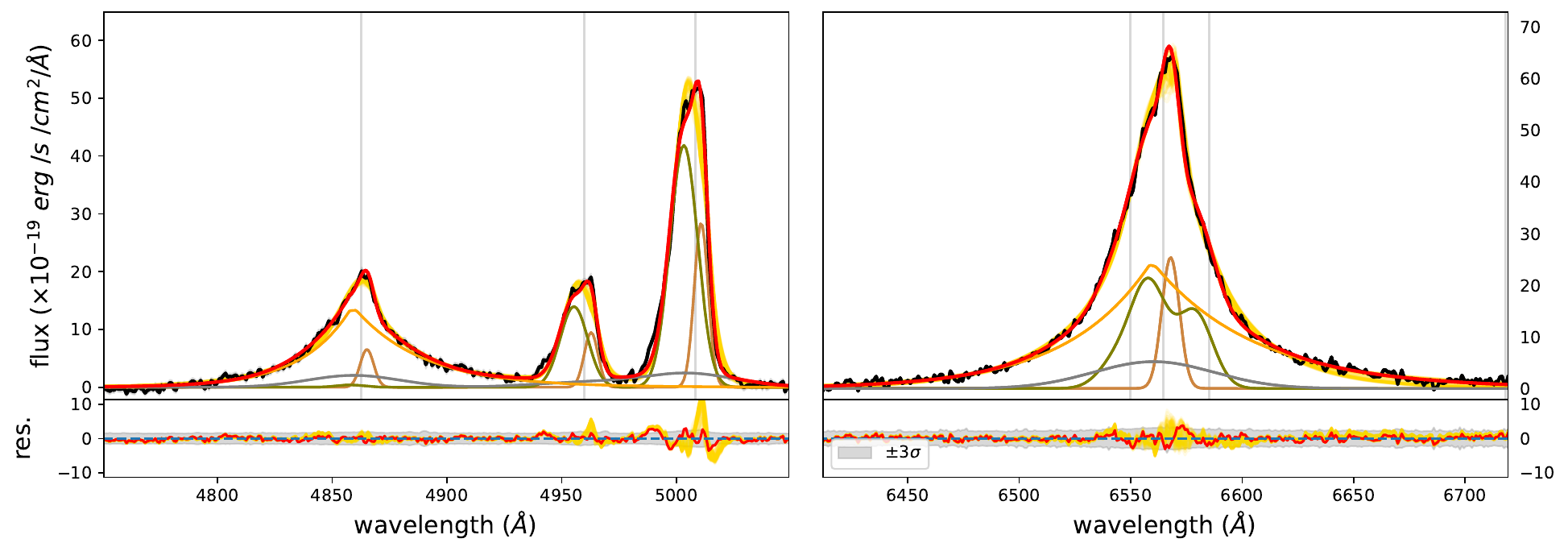}}

    \caption{Kinematic decomposition of the integrated spectrum of \HSC. The left panels show the \hb--\oiii\ region, while the right panels show the \ha--\oi--\nii\ region. The total models are selected among the 100 MC realisations, and are shown in gold, while the solution with the lowest $\chi^2$ among them is highlighted in red. 
    In the residual panels, we show the corresponding residuals.}
    \label{fig:J0859integmodel}%
    \end{figure*}

\section{The case of the LRD GN-9771: effect of a faint nuclear continuum on the sRMS}\label{App:GN9771}

\begin{figure*}
   \centering
    \includegraphics[width=0.85\textwidth]{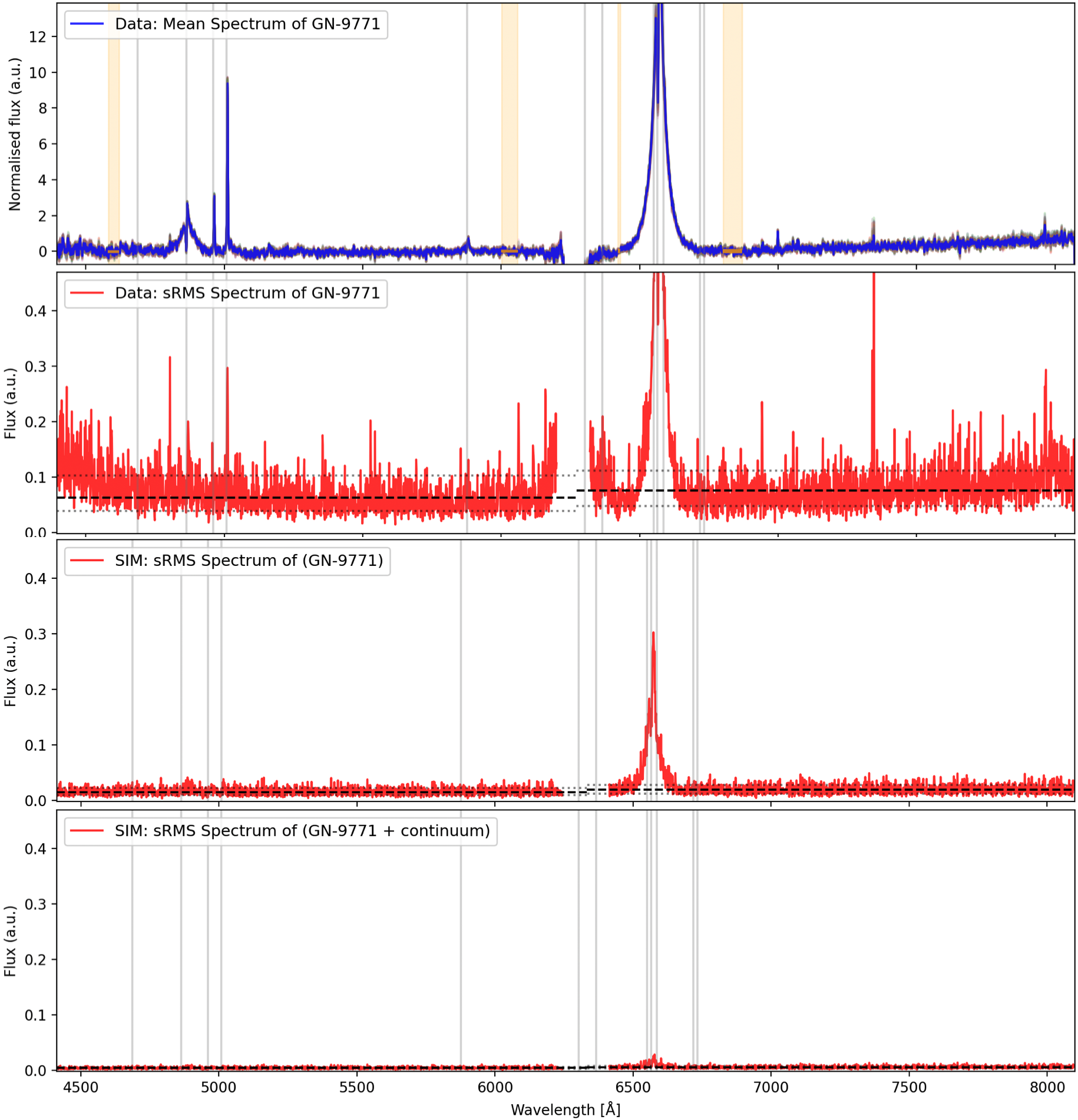}

    \caption{Mean and sRMS spectra of the $z\sim5.5$ little red dot GN-9771 and corresponding simulations. The top panel shows the mean spectrum extracted from the observed JWST/NIRSpec IFS cube, while the second panel shows the corresponding sRMS spectrum. The third panel presents the sRMS spectrum obtained from a simulated observation based on the integrated GN-9771 spectrum and the same observing configuration, illustrating the contribution of continuum fluctuations to the sRMS signal. The bottom panel shows a corresponding simulation in which the GN-9771 spectrum is combined with an additional unresolved point source with a flat continuum of $5\times10^{-19}~\mathrm{erg\,s^{-1}\,cm^{-2}\,\AA^{-1}}$. The high sRMS level in the GN-9771 simulations results from the faint continuum of the source, causing residual spectral variations after normalisation and continuum subtraction. Consequently, both narrow and broad emission-line features are retained in the sRMS spectrum, closely resembling those in the mean spectrum. In contrast, the additional continuum component substantially reduces the sRMS level, with only a weak residual feature remaining.}
    \label{fig:GN9771}%
    \end{figure*}

The applicability of the sRMS technique depends on the presence of a sufficiently strong and unresolved nuclear emission. To illustrate the effect of a faint continuum, we consider the $z\sim5.5$ LRD GN-9771 (\citealt{Matthee2024}), observed as part of the programme \#5664 (PI Jorryt Matthee). Figure~\ref{fig:GN9771} compares the sRMS obtained from the real NIRSpec IFS observations with controlled simulations based on the observed integrated spectrum.

The top panel of Fig.~\ref{fig:GN9771} shows the mean spectrum extracted from the observed data cube, while the second panel shows the corresponding sRMS spectrum. In contrast to the behaviour observed for \VDES, the sRMS of GN-9771 exhibits a high zero-level and retains spectral features associated with both narrow and broad emission components. In particular, the line features closely resemble those present in the mean spectrum, indicating that the sRMS is no longer efficiently suppressing the nuclear emission.

To investigate the origin of this behaviour, we generated a mock NIRSpec observation using the observed integrated spectrum of GN-9771 associated with a point source as input, adopting the observing configuration and exposure time of the real observations (PID~5664; PI: Matthee). The resulting sRMS is shown in the third panel of Fig.~\ref{fig:GN9771}. 
The zero-level of the simulated sRMS is determined solely by the observational noise and is lower than that measured in the real data. The higher zero-level in the real data may indicate a spatial extension of the continuum, consistent with the findings of \citet{Ishikawa2026}. Nevertheless, the point-source simulation reproduces an sRMS peak at the position of the \ha BLR, demonstrating that this feature does not require spatially extended line emission. Instead, it can arise from the treatment of the faint nuclear continuum: imperfect continuum modelling and subtraction can introduce small aperture-to-aperture variations in the residual BLR profile, which are then propagated into the sRMS and produce the observed spectral feature.

As a further test, we repeated the simulation using the same GN-9771 emission-line spectrum but adding to its faint continuum a spatially unresolved continuum of $f_\lambda = 5\times 10^{-19}$~\ergs~cm$^{-2}$~\AA$^{-1}$ at 5100$\AA$, corresponding to a continuum level $\sim20$ times higher than that of GN-9771 and similar to that measured for \VDES. The added continuum was described by a power law with $F_\lambda\propto\lambda^{-2}$, as observed in \VDES. This increases the continuum S/N to $\sim30$ in the vicinity of \ha.
The resulting sRMS spectrum, shown in the bottom panel of Fig.~\ref{fig:GN9771}, is substantially reduced in amplitude, with only a marginal residual broad feature at $\sim 3\sigma$ level. This controlled experiment confirms that the high sRMS observed for GN-9771 is primarily driven by the low S/N (and possible extension) of the continuum rather than by the intrinsic spatial extent of its emission-line components.

This example illustrates an important limitation of the technique: the sRMS is most effective for compact type~1 AGN with a strong nuclear continuum, where the individual spectra can be reliably normalised and continuum-subtracted. For LRD sources, applying the technique requires either more luminous targets and/or substantially longer integrations to achieve a sufficiently high continuum S/N.

\end{appendix}

\end{document}